\documentclass[12pt, letterpaper, onecolumn,oneside]{report}

\usepackage[left=3cm,top=4cm,right=3cm,bottom=2.5cm]{geometry} 

\usepackage{textcomp}
\usepackage{graphicx}
\usepackage{makeidx}
\usepackage{amsmath}
\usepackage{subfigure}
\usepackage{amssymb}
\usepackage{hyperref}

\hypersetup{colorlinks=false,bookmarksopen,bookmarksnumbered,citecolor=blue}

\begin{document}

\begin{titlepage}

\newpage
\begin{figure} 
\begin {center}
\includegraphics[width=0.15\textwidth]{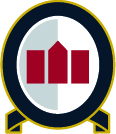}
\end {center}
\end{figure}

\begin{center}
  {  UNIVERSIDAD ANDR\'ES BELLO \\
     FACULTAD DE CIENCIAS EXACTAS \\
   DEPARTAMENTO DE CIENCIAS F\'ISICAS \\ }
\end{center}

  \bigskip

\begin{center}
\LARGE { The Geometry of 6D, $\mathcal N=(1,0)$ Superspace and its Matter Couplings }
\end{center}
\bigskip
\begin{center}
  {Thesis presented in candidacy for the degree of Master of Science}
       \end{center}
\bigskip
   \begin{center}
Author: \\
  {\sc   C\'esar Patricio Arias Huerta} \\
  \bigskip 
  \bigskip
  \bigskip
\begin{tabular}{ll}
Advisor &: {\sc Dr. William D. Linch iii } \\
Advisory Committee   &: {\sc Dr.  Rodrigo Aros}\\
                                   &: {\sc Dr.  Osvaldo Chand\'ia}\\
                                   &: {\sc Dr.  Brenno Vallilo}
\end{tabular}
\end{center}

  \vspace*{0.5 in}
  \begin{center}
   Santiago de Chile, 2014 \\
       \end{center}
\end{titlepage}

\newpage

 \begin{center}
 {\bf \Large{Abstract} }\\
\end{center}
 ~\\
 
This thesis is dedicated to the study of the geometry of six-dimensional superspace,
endowed with the minimal amount of supersymmetry. In the first part of it,  we unfold 
the main geometrical features of such superspace by solving completely the 
Bianchi identities for the constrained superspace torsion, which allow us to determine 
the full six-dimensional derivate superalgebra. Next, the conformal structure of the 
supergeometry is considered. Specifically, it is shown that the conventional torsion constraints 
remain invariant under super-Weyl transformations generated by a real scalar superfield parameter. \\
 In the second part of this work, the field content and superconformal matter couplings of the supergeometry are explored. The component field content of the Weyl multiplet is presented and the question of how this multiplet emerges in superspace is addressed. Finally, the constraints that conformal invariance imposes on some matter representations are analyzed. 

\newpage

\begin{center}
 {\bf  \Large{Acknowledgements} }
\end{center}
~\\

I am indebted to many people who made this work possible, 
and without whom I would not have been able to complete this important 
step in my career.  Above all, I would like to express my deepest gratitude
to my family, for being such an inexhaustible source of unconditional love and support,
in every aspect of my life.\\
I sincerely thank my advisor and friend, 
Professor William D. Linch {\sc iii},  whose deep understanding 
and passion for physics continuously motivated me. Thanks for 
patiently teaching me, for the constant encouragement and support,  
but most importantly, thanks for believing in me. \\
I thank to all people within the Physical Sciences Department at Universidad Andr\'es Bello, 
for having made of it such an unique place to pursue my master studies. I thank Professors Rod(rigo) Aros, Rodrigo Olea and Brenno Vallilo, from whom I have learnt great physics over the past years. I also thank Professor Osvaldo Chand\'ia for take the time of being part of the advisory committee for this thesis, and especially to Professor Per Sundell for his encouragement during the final stage of my studies.\\
Finally, my most profound appreciation goes to \emph{Kari}, for teaching me the genuine meaning of sharing life with someone.  \\

\begin{center}
\linespread{0.6}
\tableofcontents       
\linespread{2}
\end{center}

\newpage

\chapter{Introduction}
\label{Introduction}
Since the birth of modern science, the concept of symmetry has been extremely fruitful in every aspect of physics. 
It is not an accident that, each time we uncover the underlying symmetries that characterize a certain physical system, we can further understand, in a much deeper way,  such a system.

Two of the most beautiful realizations of the notion of symmetry, are the concepts of \emph{gauge symmetry} and \emph{supersymmetry}. Gauge symmetry is a remarkable symmetry simply because we can explain almost everything around us, at the fundamental level, in terms of such concept. Three of the four fundamental interactions in nature - the strong, weak, and electromagnetic interactions -  can be understood, in a unified way, in terms of a gauge theory: the standard model of particle physics.  As an outcome of this model, we know that gauge fields (bosons) mediate forces between particles described by matter fields (fermions).

Supersymmetry \cite{Golfand:1971, Volkov:1973, Wess:1974}, on the other hand, is a bizarre symmetry linking completely different type of particles. It relates bosons (force carriers) and fermions (matter building blocks), in such a way that every bosonic degree of freedom possesses a fermionic \emph{superpartner}, and vice versa.
 Although it has not been tested experimentally\footnote{Supersymmetry is not an exact symmetry.
 The fact that we have not yet found any superpartner particle implies that supersymmetry must be broken at a energy scale above what we have been able to measure. Nevertheless, the current operation of the Large Hadron Collider (LHC), the most extraordinary particle collider ever made,  holds the possibility of detecting evidence in favor of it. As of this writing, this has not happened.}, supersymmetry represents, without doubt, one of the cornerstones of modern theoretical physics. Its applications run from condensed matter and cosmology to particle phenomenology, superstring theory and mathematical physics, turning it into a central tool in the quest for our understanding of fundamental phenomena. 
 
There are several reasons to pursue the study of supersymmetric theories. First of all, the supersymmetry algebra is the unique nontrivial spacetime extension of the Poincar\'e algebra consistent with  four-dimensional quantum field theory, 
being the largest possible symmetry of the S-matrix \cite{Haag:1975}. Within the context of the minimal supersymmetric standard model, it provides a resolution of the hierarchy problem and the gauge coupling unification. In Cosmology, it also provides natural candidates for the particle spectrum of (cold) dark matter. Finally, its local version, \emph{supergravity}
 \cite{Freedman:1976, Deser:1976}, has become an entire field of research mainly because it emerges as the low energy limit of superstring theory, playing a central role in the realization of the AdS/CFT correspondence \cite{Maldacena:1998}.
 
There exist two approaches when dealing with supersymmetric theories. The most used one is the standard approach of component-fields, also known as ``tensor calculus". In this case, supersymmetry is not manifest.  The second, less used route, is the superfield \cite{Salam:1974, Ferrara:1974} or superspace formulation, in which supersymmetry is manifest. Superspace emerges as a geometrical realization of supersymmetry where supersymmetry transformations are simply translations in this space which contains, in addition to the familiar bosonic coordinates, fermionic directions. It turns out that all important concepts of differential geometry can be extended to superspace, although the description of these spaces can be quite complicated (see for instance, the standard references in the subject \cite{Gates:1983nr, Wess:1992cp, Buchbinder:1998qv}). Nevertheless, this allows for the definition and study of curved supermanifolds.

In the present work, the geometry of six-dimensional, $\mathcal N=(1,0)$ superspace is considered. Recently, superconformal models in six dimensions have captured some interest. There are at least three good reasons to focus on (1,0) superconformal models. Firstly, these models are the maximal off-shell subgroup of 
$\mathcal N=(1,1)$ and (2,0) supersymmetric formulations. This fact  allows, for instance, the enhancement of (1,0) supersymmetry to (2,0), through  the addition of a collection of (1,0) superfields (hypermultiplets) \cite{Samtleben:2011fj}.  These (2,0) theories describe the low energy limit of multiple five-branes, for which no Lagrangian description is known\footnote{Recall that, while perturbative arguments appear to rule out local, unitary QFTs in six dimensions, string theory nevertheless predicts the existence of a fully interacting such theory related to the low energy dynamics of multiple coincident five-branes \cite{Witten:1995zh}.}. Also within the context of string theory, the six-dimensional (1,0) theory appears as the target space for the covariant superstring on a K3 surface \cite{Chandia:2011su}, as well as playing a central  role in the study of the AdS$_{7}$/CFT$_6$ correspondence.

This thesis is an attempt to collect and further develop the most important results regarding the geometry of six-dimensional (1,0) superspace presented in \cite{Linch:2012zh}, and it is organized as follows: In section \ref{Supergeometry} we will solve the supergravity Bianchi identities subject to a set of conventional torsion constraints. We will elucidate, by consistency of these identities, that the full superalgebra of covariant derivatives can be written in terms of two dimension-1 superfields. Consequently, all torsions and curvatures will be expressed in terms of such fields. In section 
\ref{ConformalStructure} we will impose the invariance of the conventional constraints under super-Weyl transformations. In particular we will deduce the set of transformation rules that superfields and covariant derivatives must satisfy in order to realize the aforementioned conformal invariance. Section \ref{FieldContent} is devoted to the study of the field content of the superspace theory studied in the previous sections. The Weyl (conformal) multiplet \cite{Bergshoeff:1985mz} is reviewed and the question of how this multiplet emerges in superspace is considered. Finally, in section \ref{MatterCouplings} we investigate the constraints that super-Weyl transformations impose on matter fields. The cases of the abelian vector and tensor multiplets are studied with some detail. We conclude this work with some final comments in section \ref{ConcludingRemarks}. Notation and conventions are defined in appendix \ref{appendixA} and a supergeometry summary is presented in appendix \ref{appendixB}.


\chapter{Supergeometry}

\label{Supergeometry}
This chapter is dedicated to the study of the general structure of $\mathcal N=(1,0)$, six-dimensional superspace\footnote{Minimal supersymmetry in six dimensions (8 real supercharges) has the two different formulations, depending on the chirality of the chosen supergenerators. These are denoted by $\mathcal N=(1,0)$ and $\mathcal N=(0,1)$. Both superalgebras are isomorphic.}, suitable for a description of superfield supergravity. A superspace formulation of minimal supergravity corresponds to selecting out a specific subspace from the space of all possible supergemetries, by imposing torsion constraints. Such constraints allow us to solve the supergravity Bianchi identities that covariant derivatives must satisfy. Perhaps, the most important outcome arising from these Bianchi identities is the fact that supercurvature is, in the end, a redundant object.  More precisely, after solving the Bianchi identities one is able to express the supercurvature entirely in terms of the supertosion.\footnote{In superspace literature, this fact is known as \emph{Dragon theorem}. For a more detailed discussion see \cite{Buchbinder:1998qv}.}
Following this reasoning, we derive in detail the new six-dimensional curved superspace geometry presented in \cite{Linch:2012zh}, suitable for a superspace description of simple supergravity in six dimensions. In particular, we calculate the full six-dimensional curved superspace derivative algebra, through solving completely the Bianchi identities for the constrained supersapace torsion.
\section{The setup}
Let us consider a curved six-dimensional superspace\footnote{$\mathcal M^{p|q}$ denotes the curved supermanifold constructed with $p$ bosonic coordinates and $q$ fermionic directions. This notation makes manifest the geometrical nature of supersymmetry in superspace:
 the number of fermionic coordinates is equal to number of supercharges of the theory (eight in our case), which allow us to implement supersymmetry transformations as translations generated by each of these supercharges in such fermionic directions.} $\mathcal M^{6|8}$, parametrized  through the supercoordinates
\begin{equation}
z^{M}=(x^{m}, \theta^{\mu}_{i} ) \, \, \, , m=0, \cdots , 3; 5,6   \, , \, \,  \mu=1,2,3,4 \, , \, \,  i=\underline{1},\underline{2} \, , 
\end{equation}
with $m$ labeling bosonic coordinates $(x^m)$, and $\mu$ labeling fermionic ones $(\theta^{\mu}_{i})$. The index $i$ is 
related to the R-symmetry of the theory, as indicated below. Further details of conventions and notation are given in appendix \ref{appendixA}.\\
Choosing the structure group to be  $\text{G}=\text{SO}(5,1) \times \text{SU}(2)$, we expand the covariant derivative $\mathcal D_A=(\mathcal D_a, \mathcal D_{\alpha i} )$ as
\begin{eqnarray}
\mathcal D_A = E_A + \Omega_A + \Phi_A ~,
\end{eqnarray}
with $E_A$, $\Omega_A$ and $\Phi_A$ denoting the coframe, and the Lorentz and SU(2) connections, respectively.  Each piece can be written in terms of the generators of the superalgebra
\begin{eqnarray}
E_A= E_A{}^{M} \partial_M \, , \, \,  \Omega_A = \tfrac{1}{2}\Omega_{A}{}^{bc} M_{bc}  \, , \, \, \Phi_A = \Phi_A{}^{ij} J_{ij} ~,
\end{eqnarray}
where $\partial_M = \partial / \partial z^M$, $M_{bc} = - M_{cb}$ is the Lorentz generator and $J^{ij}=J^{ji}$ is the SU(2) R-symmetry generator. These are defined through  their action on spinor derivatives as 
\begin{equation}
\label{LRep}
[M_{ab}, \mathcal D_{\gamma k}] = -\tfrac{1}{2} (\gamma_{ab})_{\gamma}{}^{\delta} \mathcal D_{\delta k}  \, \, , \, \, 
[J^{ij}, \mathcal D_{\gamma}{}^{k}] = \varepsilon^{k(i} \mathcal D_{\gamma}{}^{j)} ~.
\end{equation}
From the spinor representation of the Lorentz generator, it also follow that
\begin{equation}
[M_{ab}, \mathcal D_c] = 2\, \eta_{c[a} \mathcal D_{b]} ~.
\end{equation}
The (anti-)commutation relations of covariant derivatives defines torsion $T_{AB}{}^{C}$, Lorentz curvature $R_{AB}{}^{cd}$,
and SU(2) field strength $F_{AB}{}^{ij}$
\begin{equation}
[\mathcal D_A, \mathcal D_B \} = T_{AB}{}^{C} \mathcal D_C + \tfrac{1}{2} R_{AB}{}^{cd} M_{cd} + F_{AB}{}^{ij} J_{ij}   ~,
\label{algebra}
\end{equation}
where we use $[\mathcal D_A, \mathcal D_B \}$ to denote a graded commutator (anti-commutator if both A and B are fermionic indices, commutator otherwise). Relations (\ref{algebra}) obey Bianchi identities
\begin{align}
[ \mathcal D_A, [ \mathcal D_B, \mathcal D_C    \}  \}  &+
(-1)^{ \varepsilon_{A} (\varepsilon_{B} + \varepsilon_{C}) }  [ \mathcal D_B, [ \mathcal D_C, \mathcal D_A    \}  \}  \nonumber \\
&+ (-1)^{\varepsilon_{C} (\varepsilon_{A} +\varepsilon_{B})}[ \mathcal D_C, [ \mathcal D_A, \mathcal D_B   \}  \}= 0  ~,
\label{BI}
\end{align}
where $\varepsilon_M$ stands for the Grassmann parity function: $\varepsilon_M=0$ if $M=m$ (bosons) and $\varepsilon_M=1$ if
$M=\mu$ (fermions).\\
In order to solve the previous identities, we need to impose conventional constraints on the torsion. These \emph{fix completely the geometry} in the sense that they isolate a specific subspace in the the space of all possible supergeometries. Such constraints are taken to be\footnote{These constraints are formally identical to those of five-dimensional conformal superspace supergravity of \cite{Kuzenko:2008wr}. }
\begin{eqnarray}
\label{cc1}
T_{\alpha i \beta j}{}^{c} = 2\varepsilon_{ij} (\gamma^c)_{\alpha \beta}  \, &\text{(dimension 0)}~,&  \\
\label{cc2}
T_{\alpha i \beta j}{}^{\gamma k} =0 \, \, , \, \,  T_{\alpha i b}{}^{c}=0  \,               &\text{(dimension $\tfrac{1}{2}$)}~,&   \\
\label{cc3}
T_{a b}{}^{c} = 0 \, \, , \, \,  T_{a\, \beta (j}{}^{\beta}{}_{k)} =0  \, &\text{(dimension 1)}~.&
\end{eqnarray}
Once the constraints (\ref{cc1})-(\ref{cc3}) are introduced, Bianchi identities (\ref{BI}) can be solved. For this purpose, it is convenient to organize the study of the identities according to the increasing mass-dimension of them. This dimensionality depends on the index combination $(A,B,C)$ that we take in (\ref{BI}). The number of possibilities for such combinations is four, and they give rise to the following set of identities \footnote{Here and through this work, we adopt the usual notation for composite indices $\underline \alpha := \alpha \, i$.}
\begin{eqnarray}
\label{sss}
 0&=& 2\, [ \mathcal D_{(\underline \alpha}, \{ \mathcal D_{\underline \beta)}, \mathcal D_{\underline\gamma }\}  ] 
+ [ \mathcal D_{\underline \gamma }, \{ \mathcal D_{\underline \alpha}, \mathcal D_{\underline \beta} \}   ]  \, \, \, \, \, \, \, \, ; \, \, \, \, \, 
(A=\underline{\alpha}, B=\underline{\beta}, C=\underline{\gamma}) ~,\\
\label{ssv}
0&=& 2\, [ \mathcal D_{(\underline \alpha}, [ \mathcal D_{\underline \beta)}, \mathcal D_{c}]   \}  
+ [ \mathcal D_{c}, \{ \mathcal D_{\underline \alpha}, \mathcal D_{\underline \beta} \}   ]   \, \, \, \, \, \, \,  \, \, \, \, ; \, \, \, \, \, 
(A=\underline{\alpha}, B=\underline{\beta}, C=c)~, \\
\label{svv}
0&=& [ \mathcal D_{\underline\alpha},  [\mathcal D_{b}, \mathcal D_{c} ]   \} 
+ 2\, [ \mathcal D_{[b},  [\mathcal D_{c]}, \mathcal D_{\underline \alpha} \}   \}    \, \, \, \, \, \, \, \, \, \, \,  \, \, \, ; \, \, \, \, \, 
(A=\underline{\alpha}, B=b, C=c)~, \\
\label{vvv}
0&=&2\, [ \mathcal D_{[a} , [\mathcal D_{b]}, \mathcal D_{c}]] , [\mathcal D_{c}, [\mathcal D_{a}, \mathcal D_{b}]] 
\, \, \, \, \, \, \,\, \, \, \, \, \, \, \, \, \,  \, \,   \, \, \, \, \, ; \, \, \, \, \, 
(A=a, B=b, C=c) ~.
\end{eqnarray}
Furthermore, within each of these four equation, there are four independent pieces: two parts proportional to the covariant derivatives (fermionics and bosonic, $\mathcal D_{\alpha i}$ and $\mathcal D_{a}$), as well a two parts proportional to the Lorentz and 
SU(2) generators, $M_{ab}$ and $J_{ij}$, respectively. Table (\ref{Table2.1}) below summarizes the splitting just described, together with the mass-dimension of each independent piece within the Bianchi identities.

\begin{table}[h]
\begin{center}
\begin{tabular}{|c|c|c|c|c|} 
\hline 
\hline 
          & $\mathcal D_{\alpha i}$ & $\mathcal D_{a}$  & $M_{ab}$& $J_{ij}$\\ 
\hline 
$[ sss\}$ & 1 & $\tfrac{1}{2}$ & $\tfrac{3}{2}$& $\tfrac{3}{2}$\\ 
\hline 
$[ ssv\}$& $\tfrac{3}{2}$ &1 &2&2\\ 
\hline 
$[ svv\}$ &2 &$\tfrac{3}{2}$&$\tfrac{5}{2}$&$\tfrac{5}{2}$\\ 
\hline 
$[ vvv\}$ & $\tfrac{5}{2}$ & 2& 3& 3\\ 
\hline 
\hline
\end{tabular}
\label{Table2.1}
\caption{\small Summary of Bianchi identities we study in this section. Here, ``$s$" stands for a spin index, and ``$v$" for a vector one. In this way, for instance, the first row give us the dimensionality of each part within the Bianchi identity (\ref{sss}), the second row indicates the dimension of each piece in (\ref{ssv}), and so forth.}
\end{center}
\end{table}
In the next sections, we proceed to solve in detail the Bianchi identities up to dimension-2. The outcome of this procedure will be the full algebra of covariant derivatives which characterizes the curved supergeometry. We will express curvatures and field-strengths completely in terms of the torsion, and we will find the constraints that the  supergravity fields entering in the algebra must satisfy.
\section{Dimension-1 Bianchi identities}
Dimension-1 identities arise by taking  the part proportional to the spinorial derivative inside the $[sss\}$-identity (\ref{sss}),  and  
the piece proportional to the vector derivative within the $[ssv\}$-identity (\ref{ssv}), as indicated in table (\ref{Table2.1}). 
As a first step, let us focus on the latter. This is given by
\begin{eqnarray}
\label{ssvv1}
0= 2i\, (\gamma^{b})_{\alpha \gamma} T_{\beta j \, a}{}^{\gamma}{}_{i}
  +2i\, (\gamma^{b})_{\beta \gamma} T_{\alpha i \, a}{}^{\gamma}{}_{j} - R_{\alpha i \beta j \, a}{}^{b}  ~.
\end{eqnarray}
From here, it is clear that we can solve for the dimension-1 curvature in terms of the dimension-1 torsion
\begin{eqnarray}
R_{\alpha i \beta j \, a}{}^{b} = 2i\, (\gamma^{b})_{\alpha \gamma} T_{\beta j \, a}{}^{\gamma}{}_{i}
  +2i\, (\gamma^{b})_{\beta \gamma} T_{\alpha i \, a}{}^{\gamma}{}_{j}  ~.
\label{ssvv2}
\end{eqnarray}
Moreover, demanding the antisymmetry of the curvature on its Lorentz indices, that is imposing $R_{\alpha i \beta j }{}^{(ab)}=0$, we get 
\begin{equation}
T^{(a}{}_{\beta}{}^{j \, \gamma k} (\gamma^{b)})_{\gamma \delta} +
T^{(a}{}_{\delta}{}^{k \, \gamma j} (\gamma^{b)})_{\gamma \beta} =0  ~.
\label{ssvv3}
\end{equation}
The above constraint on the dimension-1 torsion is particularly strong, since it implies the general form that such torsion must have. Expanding out the torsion into irreducible pieces\footnote{Note that, given the torsion expansion declared here, this theory will not contain Lorentz zero forms.}
\begin{equation}
T_{\beta}{}^{j}{}_{a}{}^{\gamma k} = A_a \, \varepsilon^{jk} \delta_{\beta}^{\gamma} 
+ B^{b} \varepsilon^{jk} (\gamma_{ab})_{\beta}{}^{\gamma}
+ C^{b\, jk} (\gamma_{ab})_{\beta}{}^{\gamma} + N_{abc}  \varepsilon^{jk} (\gamma^{bc})_{\beta}{}^{\gamma}  
+M_{abc}{}^{jk} (\gamma^{bc})_{\beta}{}^{\gamma} 
\end{equation}
and plugging this general expression back in (\ref{ssvv3}), one finds that, necessarily, the superfields 
$A_{a}$ and $B_{b}$ must vanish, as well as the tensor superfield  $M_{abc}{}^{jk}$. This means that \emph{the torsion and curvature tensors defined in} (\ref{algebra}) \emph{can be expressed entirely in terms of the dimension-1 superfields $N_{abc}$ and $C_{aij}$, and their covariant derivatives}. It also follows that these superfields must have the symmetries
\begin{equation}
N_{abc}= N_{[abc]} \, \, \, \, \, ; \, \, \, \, \, C_{a\, ij} = C_{a\, (ij)} ~.
\end{equation}
Therefore, we find that the dimension-1 torsion is defined by 
\begin{equation}
T_{\gamma k \, a}{}^{\delta l} \mathcal D_{\delta l} 
         := [\mathcal D_{\gamma k} , \mathcal D_{a}] | 
          = -C^{b}{}_{kl} (\gamma_{ab})_{\gamma}{}^{\delta} \mathcal D_{\delta}{}^{l}
   									+ N_{abc} (\gamma^{bc})_{\gamma}{}^{\delta} \mathcal D_{\delta k}~,
\label{dim1T}									
\end{equation}
and because of the (spin) traceless of the gamma 2-forms in the above commutator, we indeed have a stronger dimension-1 conventional constraint
\begin{equation}
T_{a \, \beta j}{}^{\beta k} =0  ~.
\end{equation}
Following our analysis, the dimension-$\tfrac{1}{2}$ covariant derivatives obey an anti-commutation relation which can be expanded over the superfields $C_{aij}$ and $N_{abc}$. The most general form consistent with the dimension-0 and $\tfrac{1}{2}$ torsions is
\begin{align}
\{ \mathcal D_{\alpha i}, \mathcal D_{\beta j} \} &= 2i\, \varepsilon_{ij} (\gamma^a)_{\alpha \beta} \mathcal D_a
+ia\, (\gamma^{abc})_{\alpha \beta} C_{a ij} M_{bc} +ib\, \varepsilon_{ij} (\gamma_{a})_{\alpha \beta} N^{abc}M_{bc} \nonumber \\
&+ ic\, \varepsilon_{ij} (\gamma_{a})_{\alpha \beta} \tilde N^{abc}M_{bc}
   +  id\, \varepsilon_{ij} (\gamma_{a})_{\alpha \beta} C_{a}{}^{kl} J_{kl}+ ie\, (\gamma^{abc})_{\alpha \beta} N_{abc} J_{ij}  ~,
\label{dim1com}   
\end{align}
with $a,b,c,d$ and $e$ some coefficient that must be fixed by the consistency of the dimension-1 Bianchi identities. None of these coefficient can be absorbed in the normalization of the fields since this would change the coefficient in the dimension-1 torsion. Using the expansion (\ref{dim1com}) in the $[ssv\}$-identity (\ref{ssv}) and taking the dimension-1 piece (the part proportional to the vector derivative) gives
\begin{align}
0&= \left[-2ia\, (\gamma_{c}{}^{ab})_{\alpha \beta} C_{a ij}
+ 4i\, (\gamma_{c}{}^{ab})_{\alpha \beta} C_{a ij} \right]  \mathcal D_{b}  \nonumber \\
&- \left[2i\, (b N_{c}{}^{ab} + c \tilde N_{c}{}^{ab})\varepsilon_{ij} (\gamma_a)_{\alpha \beta} 
- 8i\, \varepsilon_{ij} N_{c}{}^{ab} (\gamma_a)_{\alpha \beta}\right] \mathcal D_b   ~,
\label{ssvv4}
\end{align}
where the two lines must vanish separately. On the one hand, from the terms involving the $C$ field, it follows that $a=2$. 
On the other hand, splitting $N$ into self-dual and anti-self dual parts, the second line in (\ref{ssvv4}) implies two equations: $b+c-4=0$ and $b-c-4=0$\footnote{Note that, in principle, it is possible that $N$ have a definite duality property which would eliminate one of these equations. Nevertheless, we consider here the most general case in which $N$  does not obey any duality constraint.}, which determine the values  
$b=4$ and $c=0$. The coefficients $d$ and $e$ follow from the dimension-1 piece inside the $[sss\}$-identity. Plugging the expansion (\ref{dim1com}) into 
(\ref{sss}), and taking the part proportional to the spinorial derivative,  we get  terms of the type $C\mathcal D$ and 
$N\mathcal D$, as in (\ref{ssvv4}). For simplicity, we analyze each of these terms separately. Beginning with $C\mathcal D$, we find
\begin{eqnarray}
0=2i\, \varepsilon_{ij} [\mathcal D_{\alpha \beta}, \mathcal D_{\gamma k}] | 
- \tfrac{ia}{2}\, C_{aij} (\gamma^{abc})_{\alpha \beta} (\gamma_{bc})_{\gamma}{}^{\delta} \mathcal D_{\delta k} 
+ id \varepsilon_{ij} C_{akl} (\gamma^{a})_{\alpha \beta} \mathcal D_{\gamma}{}^{l}   + \text{c.p.} ~,
\label{ssssCD}
\end{eqnarray}
where ``c.p." stands for ``cyclic permutation" of indices. Here, the first term can be re-written using the dimension-1 torsion (\ref{dim1T}) and the identity  (\ref{6FierzHigh1}) as
\begin{eqnarray}
2i\, \varepsilon_{ij} [\mathcal D_{\alpha \beta}, \mathcal D_{\gamma k}] | = 
4i\, \varepsilon_{ij} C_{a kl} \varepsilon_{\alpha \beta \gamma \delta} (\tilde \gamma^{a})^{\delta \sigma} \mathcal D_{\sigma}{}^{l}
+2i\varepsilon_{ij} C_{a kl} (\gamma^{a})_{\alpha \beta} \mathcal D_{\gamma}{}^{l} ~.
\end{eqnarray}
In this last expression, the first term vanishes under cyclic permutation since\footnote{ Recall that the SU(2) group manifold admits a non-degenerate symplectic 2-form, namely $\varepsilon_{ij}$, which acts naturally as a ``metric" tensor on such manifold, and allow us 
to map tangent space vectors  to cotangent space elements as $X^{i}=\varepsilon^{ij} X_{j}$. The existence of this object allow us to write every rank-2 antisymmetric tensor  in terms of its trace, that is  
$T_{[ij]} =\tfrac{1}{2} \varepsilon_{ij}  \, T^{k}{}_{k}$.}. 
\begin{equation}
\label{kernell}
 \varepsilon_{ij} \varepsilon_{\alpha \beta \gamma \delta} \psi_{k} +
 \varepsilon_{jk} \varepsilon_{ \beta \gamma \alpha \delta} \psi_{i}+
 \varepsilon_{ki} \varepsilon_{\gamma \alpha \beta \delta} \psi_{j} =   
 \varepsilon_{\alpha \beta \gamma \delta} \varepsilon_{[ij}\psi_{k]} \equiv 0 ~,
\end{equation}
for any $\psi$.
Now, the second term in (\ref{ssssCD}) can be simplified by using (\ref{6FierzHigh3}) and cyclic reordering to
\begin{equation}
- \tfrac{ia}{2}\, C_{aij} (\gamma^{abc})_{\alpha \beta} (\gamma_{bc})_{\gamma}{}^{\delta} \mathcal D_{\delta k} + \text{c.p} =
-4ia \, (\gamma^{a})_{\alpha \beta} C_{ak[i} \mathcal D_{\gamma j]} + \text{c.p.}
\end{equation}
so that the second and third term in (\ref{ssssCD}) combine. That is, Eq. (\ref{ssssCD}) takes the form
\begin{eqnarray}
0= 2i\varepsilon_{ij} C_{a kl} (\gamma^{a})_{\alpha \beta} \mathcal D_{\gamma}{}^{l} 
  +i(2a +d) \, \varepsilon_{ij} C_{a kl} (\gamma^{a})_{\alpha \beta} \mathcal D_{\gamma}{}^{l}   + \text{c.p.} 
\end{eqnarray}
From this is clear that $d=-6$ (recall that $a=2$).
Next, we consider the terms of the type $N\mathcal D$ inside the dimension-1 part of the $[sss\}$-identity. This gives
\begin{align}
0&= 2i\, \varepsilon_{ij} [\mathcal D_{\alpha \beta}, \mathcal D_{\gamma k}] |
  - \tfrac{ib}{2}\, \varepsilon_{ij} N_{abc}(\gamma^{a})_{\alpha \beta} (\gamma^{bc})_{\gamma}{}^{\delta} \mathcal D_{\delta k}
  -ie\, N_{abc} (\gamma^{abc})_{\alpha \beta} \, \varepsilon_{k(i} \mathcal D_{\gamma j)} + \text{c.p.}  \nonumber \\
&=-i \left(2+\tfrac{b}{2} \right) \varepsilon_{ij} N_{abc} (\gamma^{a})_{\alpha \beta} (\gamma^{bc})_{\gamma}{}^{\delta} \mathcal D_{\delta k}
-ie\, N_{abc} (\gamma^{abc})_{\alpha \beta} \, \varepsilon_{k(i} \mathcal D_{\gamma j)} + \text{c.p.}  
\end{align}
This time, the second term rearranges under cyclic permutation as
\begin{equation}
-ie\, N_{abc} (\gamma^{abc})_{\alpha \beta} \, \varepsilon_{k(i} \mathcal D_{\gamma j)} + \text{c.p.} 
= ie\, \varepsilon_{ij} N_{abc} (\gamma^{abc})_{\gamma [ \alpha} \mathcal D_{\beta] k} + \text{c.p.} 
\end{equation}
Plugging the identity (\ref{BigFierz}) and using the relation (\ref{kernell}) we obtain
\begin{equation}
2+\tfrac{b}{2} + \tfrac{3e}{2} =0 ~.
\end{equation}
Therefore, substituting our previous result $b=4$, we get the value  $e=-\frac{8}{3}$,
which fixes all the coefficients in the dimension-1 anti-commutator (\ref{dim1com}). We conclude
that
\begin{align}
\{ \mathcal D_{\alpha i}, \mathcal D_{\beta j} \} = 2i\, \varepsilon_{ij} (\gamma^a)_{\alpha \beta} \mathcal D_a
&+2i\, (\gamma^{abc})_{\alpha \beta} C_{a ij} M_{bc} +4i\, \varepsilon_{ij} (\gamma_{a})_{\alpha \beta} N^{abc}M_{bc} \nonumber \\
&-6i\, \varepsilon_{ij} (\gamma_{a})_{\alpha \beta} C_{a}{}^{kl} J_{kl}-\tfrac{8i}{3} \, (\gamma^{abc})_{\alpha \beta} N_{abc} J_{ij} ~.
\end{align}
This calculation completes the analysis of the dimension-1 identities.

\section{Dimension-$\tfrac{3}{2}$ Bianchi identities}
There are four pieces of the Bianchi identities with dimension-$\tfrac{3}{2}$, as we can read off from the table (\ref{Table2.1}). None of these is trivially fulfilled. In this section, we will analyze these four parts separately. From this analysis, we will be able to express the dimension-$\tfrac{3}{2}$ curvature, torsion and isospin field strength in terms of irreducible pieces. We will also show that Bianchi identities impose constraints on the supergravity fields $C$ and $N$, and we will find such constraints. 
\paragraph{Dimension-$\tfrac{3}{2}$ curvature} At dimension-$\tfrac{3}{2}$ level, we can write the Lorentz curvature in terms of the torsion. In order to do this, we take the part proportional to the vector derivative $\mathcal D_{a}$ of the $[svv\}$-identity (\ref{svv}). This gives
\begin{equation}
\label{svvv1}
R_{\gamma k [c\,a] b} = i\, T_{c\, a}{}^{\delta}{}_{k} (\gamma_{b})_{\delta \gamma} ~.
\end{equation}
Adding to this the signed permutation $(cab+bca-abc)$ and using the antisymmetry of $R$ on its Lorentz indices, we derive that
\begin{eqnarray}
\label{svvv2}
R_{\gamma k c\, ab} = -i\, T_{a\, b}{}^{\delta}{}_{k} (\gamma_{c})_{\delta \gamma} 
                                       +2 i\, T_{c\, [a}{}^{\delta}{}_{k} (\gamma_{b]})_{\delta \gamma} ~,
\label{dim3/2R}
\end{eqnarray} 
and thus we have an equation for the curvature in terms of the torsion.

\paragraph{Dimension-$\tfrac{3}{2}$ isospin field strength}
From the part proportional to the spinorial derivative  of the $[ssv\}$-identity (\ref{ssv}), we can get a general expression for the isospin field strength. Although this expression will depend explicitly on the torsion and curvature, it will be enough to write the field strength in terms of irreducibles. The $\mathcal D_{\gamma k}$-part of (\ref{ssv}) is given by
\begin{align}
\label{ssvs1}
0=&-2i\, \varepsilon_{ij} (\gamma^{d})_{\alpha \beta} T_{dc}{}^{\gamma}{}_{k}
+ (\mathcal D_{\alpha i} \mathbf{T}_{\beta j\, c}{}^{\gamma}{}_{k} +\mathcal D_{\beta j} \mathbf{T}_{\alpha i\, c}{}^{\gamma}{}_{k})
+ (\varepsilon_{ki} R_{\beta j \, c\, \alpha}{}^{\gamma} +\varepsilon_{kj} R_{\alpha i \, c\, \beta}{}^{\gamma}) \nonumber\\
&- (\delta_{\alpha}^{\gamma} F_{\beta j\, c\, ik} + \delta_{\beta}^{\gamma} F_{\alpha i\, c\, jk}) ~.
\end{align}
Here, we use bold font to indicate that the tensor in question is known in terms of the fields $C$ and $N$. In this case
\begin{equation}
\mathbf{T}_{\alpha i \, c}{}^{\gamma}{}_{k} := C^{d}{}_{ik} (\gamma_{cd})_{\alpha}{}^{\gamma}
            							 - \varepsilon_{ik} N_{cab} (\gamma^{ab})_{\alpha}{}^{\gamma} ~.
\end{equation}
Now, taking the trace $\alpha=\gamma$ over  (\ref{ssvs1}) and noting that $\mathbf{T}_{\gamma i \, c}{}^{\gamma k}=0$, gives
\begin{eqnarray}
\label{ssvs2}
F_{\alpha j \, c\, ik} + 4\,F_{\alpha i \, c\, jk} = 2i\, \varepsilon_{ij} (\gamma^{d})_{\alpha \beta} T_{cd}{}^{\beta}{}_{k}
     + \mathcal D_{\beta j} \mathbf{T}_{\alpha i\, c}{}^{\beta}{}_{k}  + \varepsilon_{ki} R_{\beta j \, c\, \alpha}{}^{\beta} ~.
\end{eqnarray}
We can solve for $F$ by adding to the previous equation the same expression with a factor of $-\tfrac{1}{4}$, getting 
\begin{align}
\label{ssvs3}
F_{\alpha i \, c\, jk} = \tfrac{2i}{3} \, \varepsilon_{ij} (\gamma^{d})_{\alpha \beta} T_{cd}{}^{\beta}{}_{k}
        &+ \tfrac{4}{15} \left( \mathcal D_{\beta j} \mathbf{T}_{\alpha i\, c}{}^{\beta}{}_{k} 
                                -\tfrac{1}{4}  \mathcal D_{\beta i} \mathbf{T}_{\alpha j\, c}{}^{\beta}{}_{k} \right)   \nonumber \\
        &+\tfrac{4}{15} \left( \varepsilon_{ki} R_{\beta j \, c\, \alpha}{}^{\beta} 
                               -\tfrac{1}{4}\varepsilon_{kj} R_{\beta i \, c\, \alpha}{}^{\beta}   \right)    ~.                   
\end{align}
The resulting expression for the field strength must by symmetric in its isospin indices $_{(jk)}$. Imposing such symmetry we obtain
\begin{eqnarray}
\label{ssvs4}
0= 2\,(\gamma^{d})_{\alpha \beta} T_{cd}{}^{\beta}{}_{i} - \tfrac{1}{4} (\gamma_{c}{}^{ab})_{\alpha\beta} T_{ab}{}^{\beta}{}_{i}
     - \mathbf{T}_{\alpha i \, c} ~,
\end{eqnarray}
where we have defined 
\begin{eqnarray}
 \mathbf{T}_{\alpha i \, c} := -2i\, \mathcal D_{\beta k} \mathbf{T}_{\alpha i \, c}{}^{\beta k} 
                                              + \tfrac{i}{2} \, \mathcal D_{\beta i} \mathbf{T}_{\alpha i \, k}{}^{\beta k} 
                                              = 2i\, (\gamma_{cd})_{\alpha}{}^{\beta} \mathcal D_{\beta}{}^{j} C^{d}{}_{ij}
                                                -i\, (\gamma^{ab})_{\alpha}{}^{\beta} \mathcal D_{\beta i} N_{abc}  ~. ~~~~~~~~
\end{eqnarray}
\label{ssvs5}
Contracting (\ref{ssvs4}) with $(\tilde \gamma^{c})^{\gamma \alpha}$, we can isolate the term
\begin{eqnarray}
(\gamma^{ab})_{\beta}{}^{\gamma} T_{ab}{}^{\beta}{}_{i} =\tfrac{1}{3} (\tilde \gamma^{c}) \mathbf{T}_{\alpha i \, c}
      = -\tfrac{10i}{3}\,(\tilde \gamma_{c})^{\gamma \beta} \mathcal D_{\beta}{}^{j} C^{c}{}_{ij}
      -\tfrac{i}{3}(\tilde \gamma^{abc})^{\gamma \beta} \mathcal D_{\beta i} N_{abc} ~.
\end{eqnarray}
Plugging this back in (\ref{ssvs4}), we find
\begin{equation}
\label{ssvs6}
(\gamma^{d})_{\alpha \beta} T_{cd}{}^{\beta}{}_{i} = -\tfrac{2}{3} \left[ \delta_{\alpha}^{\gamma} \delta_{c}^{d}
                             + \tfrac{1}{12} (\gamma_c \tilde \gamma^d)_{\alpha}{}^{\gamma}   \right]  \mathbf{T}_{\gamma i \, d} ~.
\end{equation}
With this we can simplify the trace of the Lorentz curvature in Eq. (\ref{svvv2}) to
\begin{equation}
\label{ssvs7}
R_{\beta i\, c \, \alpha}{}^{\beta} = \tfrac{7i}{3} \,\mathbf{T}_{\alpha i \, c}
       + \tfrac{5i}{18}\, (\gamma_c \tilde \gamma^d)_{\alpha}{}^{\beta}  \mathbf{T}_{\beta i \, d}  ~.
\end{equation}
Therefore, plugging the previous trace into (\ref{ssvs3}), we solve for the dimension-$\tfrac{3}{2}$ isospin field strength 
\begin{align}
F_{\alpha i \, c\, jk} = \tfrac{2i}{3} \, \varepsilon_{i(j} (\gamma^{d})_{\alpha \beta} T_{cd}{}^{\beta}{}_{k)}
      &-\tfrac{4}{15} \, \varepsilon_{i(j} R_{\beta k) \, c\, \alpha}{}^{\beta}  \nonumber \\
      &+ \tfrac{4}{15} \left[ \mathcal D_{\beta (j|} \mathbf{T}_{\alpha i\, c}{}^{\beta}{}_{|k)} 
                                -\tfrac{1}{4}  \mathcal D_{\beta i} \mathbf{T}_{\alpha (j\, c}{}^{\beta}{}_{k)} \right]  ~.
\label{dim3/2F}
\end{align}

\paragraph{Dimension-$\tfrac{3}{2}$ torsion}  Next, we focus on the torsion. For this, we go back to the identity (\ref{ssvs1}).  Performing the contraction with $(\gamma_{ab})_{\gamma}{}^{\delta}$ isolates the term 
$(\gamma_{ab})_{\alpha}{}^{\beta} F_{\beta j\, c\, ik}$ which must be symmetric in $_{(ik)}$. Enforcing this condition gives
\begin{align}
\label{ssvs8}
0&=(\gamma_{ab})_{\gamma}{}^{\beta} (\mathcal D_{\alpha i} \mathbf{T}_{\beta j\, c}{}^{\gamma i}
     + \mathcal D_{\beta j} \mathbf{T}_{\alpha i \, c}{}^{\gamma i}) - i\, (\gamma_{ab}{}^{d})_{\alpha \gamma} T_{dc}{}^{\gamma}{}_{j}
     +\tfrac{i}{4} \varepsilon_{ab}{}^{fgde} (\gamma_{cfg})_{\alpha \delta} T_{de}{}^{\delta}{}_{j} \cr
  &+ \tfrac{i}{2}\varepsilon_{abc}{}^{def} (\gamma_{f})_{\alpha \delta}T_{de}{}^{\delta}{}_{j} 
     -2i\, (\gamma_{c[a}{}^{d})_{\alpha \delta} T_{b] d}{}^{\delta}{}_{j} 
     -2i\, \eta_{c[a} (\gamma^d)_{\alpha \delta} T_{b] d}{}^{\delta}{}_{j}
     +3i\, (\gamma_c)_{\alpha \delta} T_{ab}{}^{\delta}{}_{j} . 
\end{align}
Contracting again with $(\tilde \gamma^c)^{\delta \alpha}$, we obtain
\begin{eqnarray}
\label{ssvs9}
0=-20i \, T_{ab}{}^{\delta}{}_{j} - 8i\, (\gamma_{[a}{}^{c})_{\gamma}{}^{\delta} T_{b] c}{}^{\gamma}{}_{j} 
   + (\tilde \gamma^c)^{\delta \alpha}(\gamma_{ab})_{\gamma}{}^{\beta} (\mathcal D_{\alpha i} \mathbf{T}_{\beta j\, c}{}^{\gamma i}
     + \mathcal D_{\beta j} \mathbf{T}_{\alpha i \, c}{}^{\gamma i})~.
\end{eqnarray}
On the other hand, contracting (\ref{ssvs6}) with $(\tilde \gamma_{b})^{\gamma \alpha}$ and anti-symmetrizing the resulting expression gives
\begin{eqnarray}
(\gamma_{[b}{}^{d})_{\beta}{}^{\gamma} T_{c] d}{}^{\beta}{}_{j} = T_{bc}{}^{\gamma}{}_{j} 
      + \tfrac{5}{9}\, (\tilde \gamma_{[b})^{\gamma \alpha} \mathbf{T}_{\alpha j \, c]}
      + \tfrac{1}{18} (\tilde \gamma_{bc}{}^{d})^{\gamma \delta} \mathbf{T}_{\delta j \, d} ~,
\end{eqnarray}
which can be plugged back in (\ref{ssvs9}) to obtain
\begin{align}
\label{ssvs10}
0=-28i\,T_{ab}{}^{\delta}{}_{j} -\tfrac{40i}{9}\,  (\tilde \gamma_{[a})^{\delta \beta} \mathbf{T}_{\beta j \, b]}
  &-\tfrac{4i}{9}\, (\tilde \gamma_{ab}{}^{c})^{\delta \beta} \mathbf{T}_{\beta j \, c}  \cr
  &+ (\tilde \gamma^c)^{\delta \alpha}(\gamma_{ab})_{\gamma}{}^{\beta} (\mathcal D_{\alpha i} \mathbf{T}_{\beta j\, c}{}^{\gamma i}
     + \mathcal D_{\beta j} \mathbf{T}_{\alpha i \, c}{}^{\gamma i}) ~.~
\end{align} 
Here, let us compute each of the four last terms independently. The second term is proportional to
\begin{align}
\label{ssvs11}
(\tilde \gamma_{[a})^{\delta \beta} \mathbf{T}_{\beta j \, b]} &= 
                   2i\, (\tilde \gamma_{abc})^{\delta \alpha}\mathcal D_{\alpha}{}^{i} C^{c}{}_{ij} 
	       - 2i\, (\tilde \gamma_{[a}) ^{\delta \alpha}\mathcal D_{\alpha}{}^{i} C_{b] ij}  \nonumber\\
	      &-i\, (\tilde \gamma_{[a}{}^{cd}) ^{\delta \alpha}\mathcal D_{\alpha j} N_{b] cd}
	       -2i\, (\tilde \gamma^{c})^{\delta \alpha} \mathcal D_{\alpha j} N_{abc} ~,
\end{align}
while the third is given by
\begin{align}
\label{ssvs12}
(\tilde \gamma_{ab}{}^{c})^{\delta \beta} \mathbf{T}_{\beta j \, c}&=
     -6i\, (\tilde \gamma_{abc})^{\delta \alpha}\mathcal D_{\alpha}{}^{i} C^{c}{}_{ij} 
	       +16i\, (\tilde \gamma_{[a}) ^{\delta \alpha}\mathcal D_{\alpha}{}^{i} C_{b] ij}  \cr
    &+4i\, (\tilde \gamma_{[a}{}^{cd})^{\delta \alpha} \mathcal D_{\alpha j} N_{b] cd}
    + 2i\, (\tilde \gamma^{c})^{\delta \alpha} \mathcal D_{\alpha j} N_{abc}	       
      + 6i\, (\tilde \gamma^{c})^{\delta \alpha} \mathcal D_{\alpha j} \tilde N_{abc} ~. ~
\end{align}
The last two terms in (\ref{ssvs10}) expand out to give
\begin{align}
\label{ssvs13}
 (\tilde \gamma^c)^{\delta \alpha}(\gamma_{ab})_{\gamma}{}^{\beta} (\mathcal D_{\alpha i} \mathbf{T}_{\beta j\, c}{}^{\gamma i}
     + \mathcal D_{\beta j} \mathbf{T}_{\alpha i \, c}{}^{\gamma i}) &=
     8\,(\tilde \gamma_{[a}) ^{\delta \alpha}\mathcal D_{\alpha}{}^{i} C_{b] ij}
     -8\, (\tilde \gamma^{c})^{\delta \alpha} \mathcal D_{\alpha j} N_{abc}	  \nonumber \\
  &+ 2\, (\tilde \gamma^{cde})(\gamma_{ab})^{\delta \alpha} \mathcal D_{\alpha j} N_{cde} \nonumber \\
     = 8\,(\tilde \gamma_{[a}) ^{\delta \alpha}\mathcal D_{\alpha}{}^{i} C_{b] ij}
     -8\, (\tilde \gamma^{c})^{\delta \alpha} \mathcal D_{\alpha j} N_{abc}	&-
     12\, (\tilde \gamma^{c})^{\delta \alpha} \mathcal D_{\alpha j} (N_{abc} + \tilde N_{abc}) \nonumber \\
     &+6\, (\tilde \gamma_{[a}{}^{cd}) \mathcal D_{\alpha j} (N_{b] cd} + \tilde N_{b] cd}) ~.
\end{align}
Putting all these results together, that is, replacing (\ref{ssvs11}), (\ref{ssvs12}) and (\ref{ssvs13}) in (\ref{ssvs10}), we can finally solve for the dimension-$\tfrac{3}{2}$ isospin field strength:
\begin{align}
\label{dim3/2T}
T_{ab}{}^{\gamma}{}_{k} &= -\tfrac{2i}{9}\, (\tilde \gamma_{[a}) ^{\gamma \beta}\mathcal D_{\beta}{}^{l} C_{b] kl}
   -\tfrac{2i}{9} \, (\tilde \gamma_{abc})^{\gamma \beta} \mathcal D_{\beta}{}^{l} C^{c}_{kl}
   + i\, (\tilde \gamma^{c})^{\gamma \beta} \mathcal D_{\beta k} N_{abc}   \nonumber \\
&+\tfrac{i}{3}\, (\tilde \gamma^{c})^{\gamma \beta} \mathcal D_{\beta k} \tilde N_{abc}
  -\tfrac{5i}{42}\, (\tilde \gamma_{[a}{}^{cd})^{\gamma \beta} \mathcal D_{\beta k} N_{b] cd}
   -\tfrac{3i}{14}\, (\tilde \gamma_{[a}{}^{cd})^{\gamma \beta} \mathcal D_{\beta k} \tilde N_{b] cd} ~.
\end{align}
At this point, we have studied two of the four dimension-$\tfrac{3}{2}$ identities; the $[svv\}|_{v}$ and $[ssv\}|_{s}$ pieces. The remaining two parts will give rise to the constraints on the supergravity fields $C$ and $N$. Recall that these superfields define the dimension-1 torsion according to (\ref{dim1T}).

\paragraph{Dimension-1 torsion constraints}
As we mentioned at the beginning of this section, the dimension-$\tfrac{3}{2}$ identities impose constraints on the supergravity fields $C$ and $N$. In what follows, we will show that these constraints are given by
\begin{align}
\label{constraint1}
\mathcal D_{\gamma (k} C_{\alpha\beta ij)} &=
       -\tfrac{1}{3} \varepsilon_{\alpha\beta\gamma\delta} \mathcal D_{\epsilon (k} C^{\delta \epsilon}{}_{ij)}  ~,   \\
\label{constraint2}
\mathcal D_{(\alpha i}N_{\beta) \gamma} &= -\tfrac{1}{2} \mathcal D_{\gamma i} N_{\alpha \beta}     ~,
\end{align}
and
\begin{eqnarray}
\label{constraint3}
\mathcal D_{(\alpha i} N_{\beta)\gamma}= -\tfrac{1}{4} \mathcal D_{(\alpha}{}^{j} C_{\beta) \gamma ij} ~.
\end{eqnarray}
The last equation implies
\begin{equation}
\label{constraint3.1}
\tfrac{2}{3} (\gamma^{abc} \tilde \gamma_d)_{\alpha}{}^{\beta} \mathcal D_{\beta i} N_{abc} =
\tau_{d \, \alpha}^{c \,  \gamma} (5,1) \mathcal D_{\gamma}{}^{j} C_{cij} =
-4 (\tilde \gamma_d)^{\beta\gamma} \mathcal D_{\beta j} N_{\gamma \alpha} ~.
\end{equation}
with the tensor $\tau$ defined in Eq. (\ref{tau}) below. \\
In order to derive the constraint (\ref{constraint1}), we use the part proportional to the Lorentz generator $M$ within the $[sss\}$-identity (\ref{sss}). This has the form
\begin{eqnarray}
0= 2i\, \varepsilon_{ij} (\gamma^{c})_{\alpha \beta} \left[  \tfrac{1}{2}\,R_{\gamma k \, c}{}^{ab}  
 										     + 2\, \mathcal D_{\gamma k} N_{c}{}^{ab}  \right]  \, M_{ab}
  - 2i\, (\mathcal D_{\gamma k} C_{c ij})  (\gamma^{abc})_{\alpha \beta}M_{ab}   + \text{c.p.} 
\end{eqnarray}
Completely symmetrizing all three isospin indices implies
\begin{eqnarray}
0=\mathcal D_{\gamma(k} C_{cij)}  (\gamma^{abc})_{\alpha \beta}
    + 2\,\mathcal D_{(\alpha (k} C_{cij)} (\gamma ^{abc})_{\beta) \gamma} ~.
\end{eqnarray}
Contracting this last equation with $(\tilde \gamma_d)^{\delta\beta}$ we get
\begin{eqnarray}
\label{2.59}
\left[  5\, \delta_{d}^{c} \delta_{\alpha}^{\beta} + (\gamma_{d}{}^{c})_{\alpha}{}^{\beta} \right] \mathcal D_{\beta (k} C_{c\,ij)} =0 ~.
\end{eqnarray}
In this last expression, the tensor structure
\begin{eqnarray}
\label{tau}
\tau_{a \, \alpha}^{b \, \beta} (5,1) := 5\, \delta_{a}^{b} \delta_{\alpha}^{\beta} + (\gamma_{a}{}^{b})_{\alpha}{}^{\beta}
\end{eqnarray}
is not invertible.\footnote{In general, the multiplication of these tensors is given by
\begin{eqnarray}
\tau(m,n) \tau(p,q) = \tau (mp + 5nq, (m-4n)q + np)~.
\end{eqnarray}
When $\tfrac{m}{n} = 5$ this gives $\tau(m,n) \tau(p,q) = \tau(m(p+q), n(p+q))$, implying that we can not choose the coefficients to give
$\tau(1,0)$.
}
This implies that the totally symmetric term in (\ref{2.59}) is proportional to a gamma matrix
\begin{eqnarray}
\mathcal D_{\gamma (k} C_{\alpha\beta ij)} = (\gamma_{a})_{\gamma \delta} \mathcal C^{\delta}_{ijk}   \, \, \, \, \, ;   \, \, \, \, \,
\text{with} \, \, \, \, \,   \mathcal C^{\delta}_{ijk}:= -\tfrac{1}{6} (\tilde \gamma^{b})^{\delta \beta} \mathcal D_{\beta (k} C_{b ij)} ~.
\end{eqnarray}
This last expression is equivalent to (\ref{constraint1}).\\
In order to derive the second constraint (\ref{constraint2}), it is enough to study the part proportional to the SU(2) generator inside the 
$[sss\}$-identity (\ref{sss}). We get the following expression
\begin{eqnarray}
\label{sssJ1}
0=-2i\, \varepsilon_{ij} (\gamma^d)_{\alpha \beta} \left( F_{\gamma k\, d}{}{}^{mn}-3\, \mathcal D_{\gamma k} C_{d}{}^{mn}  \right)
    + 16i\, \mathcal D_{\gamma k} N_{\alpha \beta} \, \delta_{i}^{(m} \delta_{j}^{n)}  + \text{c.p.} 
\end{eqnarray}
Contracting with $\delta_{m}^{i} \delta_{n}^{j}$ gives
\begin{eqnarray}
\label{sssJ2}
0= 48i\, \mathcal D_{\gamma k} N_{\alpha \beta} + 48i\, \mathcal D_{(\alpha k} N_{\beta) \gamma} 
     - 4i\, (\gamma^d)_{\gamma(\alpha} \left[ F_{\beta) j\, d}{}^{j}{}_{k} + 3\, \mathcal D_{\beta)}{}^{j} C_{djk}  \right] ~.
\end{eqnarray}
Using the identity (\ref{3gamma}), it follows that the first and second terms in this equation are related through 
\begin{eqnarray}
\label{sssJ3}
(\gamma^d)_{\gamma (\alpha} (\gamma^{bc})_{\beta)}{}^{\delta} \mathcal D_{\delta k} N_{dbc} =
      2\,\mathcal D_{\gamma k} N_{\alpha \beta} - 2\,\mathcal D_{(\alpha k} N_{\beta)\gamma} ~.
\end{eqnarray}
The third term in (\ref{sssJ2}) contains the (isospin) trace of the field strength. Such a term can be written in terms of derivatives of the superfields $C$ and $N$ by taking the trace of Eq. (\ref{ssvs3}) and using the trace of the Lorentz curvature (\ref{ssvs7}). This gives
\begin{eqnarray}
\label{sssJ4}
 - 4i\, (\gamma^d)_{\gamma(\alpha} F_{\beta) j\, d}{}^{j}{}_{k} = 12i\, \mathcal D_{(\alpha}{}^{j} C_{\beta) \gamma \, jk}
                       +8i\,(\gamma^d)_{\gamma (\alpha} (\gamma_{bc})_{\beta)}{}^{\delta} \mathcal D_{\delta k} N_{dbc}~.
\end{eqnarray}
Replacing (\ref{sssJ3}) and (\ref{sssJ4}) in (\ref{sssJ2}) we get
\begin{eqnarray}
\label{sssJ5}
0=-3\, \mathcal D_{(\alpha}{}^{j} C_{\beta)\gamma\, ij}+ 4\,\mathcal D_{(\alpha i} N_{\beta)\gamma}
      + 8\, \mathcal D_{\gamma k} N_{\alpha \beta}~.
\end{eqnarray}
We can now simplify this result by symmetrizing on $_{(\beta \gamma)}$, obtaining
\begin{eqnarray}
\label{sssJ6}
0=-3\, \mathcal D_{(\beta}{}^{j} C_{\gamma)\alpha\, ij}+ 4\,\mathcal D_{\alpha i} N_{\beta\gamma}
      + 20\, \mathcal D_{(\beta i} N_{\gamma)\alpha}~.
\end{eqnarray}
Finally, manipulating indices and subtracting from (\ref{sssJ5}) we get
\begin{eqnarray}
\mathcal D_{(\alpha i} N_{\beta)\gamma} = -\tfrac{1}{2} \mathcal D_{\gamma i} N_{\alpha \beta} ~,
\end{eqnarray}
which is the constraint (\ref{constraint2}). Plugging this expression back into (\ref{sssJ5}) we find the constraint (\ref{constraint3}).

\paragraph{Irreducible decomposition}
Once the dimension-1 torsion constraints have been obtained, we may expand the derivative of the fields in their Lorentz- and isospin-irreducible components, in the following way
\begin{align}
\mathcal D_{\gamma k} C_{a\, ij} &=: \mathcal C_{a \, \gamma k \, ij} + (\gamma_{a})_{\gamma \delta} \, \mathcal C^{\delta}{}_{ijk} 
            + \varepsilon_{k(i} \,\mathcal C_{a \gamma j)}+ \varepsilon_{k(i}(\gamma_{a})_{\gamma \delta} \, \mathcal C^{\delta}{}_{j)}~, \\
\mathcal D_{\gamma k} N_{\alpha \beta} &=:
  \mathcal N_{\gamma k \, \alpha \beta} + \check{\mathcal N}_{\gamma k \, \alpha \beta}  ~,\\ 
\mathcal D_{\gamma k} N^{\alpha \beta} &=: \mathcal N_{\gamma k }{}^{ \alpha \beta} 
  						+ \delta_{\gamma}^{(\alpha} \mathcal N^{\beta)}{}_{k} ~.
 \end{align}
 Under this decomposition, the content of the constraints is given by
\begin{align}
 \mathcal C_{a \, \gamma k \, ij} &= 0 ~,\\
 \mathcal C^{\delta}{}_{ijk} &=-\tfrac{1}{6}\,(\tilde\gamma^{b})^{\delta\beta} \mathcal D_{\beta(k} C_{b\, ij)} ~,   \\
 \mathcal C_{a \beta j} &=  \tfrac{1}{9}\, \tau_{a\,\beta}^{c\,\gamma} (5,1)\, \mathcal D_{\gamma}{}^{j} C_{a\, ij}     ~,    \\
 \mathcal C^{\gamma k} &=-\tfrac{1}{9} \mathcal D_{\delta l} C^{\delta \gamma\, lk} ~.
 \end{align} 
 and
 \begin{align}
 ~~~~~~~~~~~~~~~  \mathcal N_{\gamma k \, \alpha \beta}&=0 ~,\\
  \label{NpropC}
   \check{\mathcal N}_{\gamma k \, \alpha \beta} &= \tfrac{1}{2} \, \mathcal D_{(\alpha}{}^{j} C_{\beta) \gamma\, ij}  
      = -\tfrac{3}{4}\,(\gamma^{a})_{\gamma(\alpha} \, \mathcal C_{a\beta) k} ~,\\
    \mathcal N_{\gamma k }{}^{ \alpha \beta} &= \mathcal D_{\gamma k} N^{\alpha \beta} 
         							               - \tfrac{2}{5}\, \delta_{\gamma}^{(\alpha} \mathcal D_{\delta k} N^{\beta) \delta}  ~,\\
    \mathcal N^{\alpha i} &=\tfrac{2}{5} \, \mathcal D_{\beta}{}^{i} N^{\beta \alpha} ~.
\end{align}
Let us focus now on the irreducible decomposition of the dimension-$\tfrac{3}{2}$ curvature, torsion and field strength. From (\ref{dim3/2R}), we  note that the curvature is most conveniently expressed in terms of the torsion, so that we do not consider its decomposition. For the torsion 
(\ref{dim3/2T}), we expand into its irreducible pieces:
\begin{eqnarray}
T_{ab}{}^{\gamma k} =\mathfrak T_{ab}{}^{\gamma k} + (\tilde \gamma_{[a})^{\gamma \delta}\, \mathfrak T_{b] \delta}{}^{k}
                                       + (\gamma_{ab})_{\delta}{}^{\gamma} \, \mathfrak T^{\delta k} ~,
\end{eqnarray}
under which we find
\begin{align}
\mathfrak T_{ab}{}^{\gamma k} &=
           \tfrac{i}{21}\, (\tilde \gamma_{[a}{}^{cd})^{\gamma \delta} \mathcal D_{\delta}{}^{k} N^{(-)}_{b] cd}
           -\tfrac{2i}{5}\, (\tilde \gamma_{[a}{}^{cd})^{\gamma \delta} \mathcal D_{\delta}{}^{k} N^{(+)}_{b] cd}
          +\tfrac{6i}{5}\, (\tilde \gamma^{c})^{\gamma \delta} \mathcal D_{\delta}{}^{k} N^{(+)}_{abc}   ~,  \\
\mathfrak T_{a\, \beta j}&=
          -\tfrac{i}{9}\, \tau_{a\, \beta}^{c\, \gamma} (5,1) \mathcal D_{\gamma}{}^{i} C_{c\, ij} 
          -\tfrac{i}{3}\,  (\tilde \gamma_{a})^{\gamma \delta} \mathcal D_{\delta j} N_{\beta \gamma}  
          =-i\, \mathcal C_{a\, \beta j} + \tfrac{i}{3}\,(\tilde \gamma_{a})^{\gamma \delta}   \check{\mathcal N}_{\gamma j \, \delta \beta} ~,      \\
\mathfrak T^{\delta k} &=  \tfrac{i}{9}\,\mathcal D_{\gamma l}C^{\gamma \delta\, lk} 
                  			    + \tfrac{i}{15}\, \mathcal D_{\gamma}{}^{k} N^{\gamma \delta}
			    = -i\, \mathcal C^{\delta k} + \tfrac{i}{6}\, \mathcal N^{\delta k} ~.
\end{align}
Here, the first term in $\mathfrak{T}_{ab}{}^{\gamma k}$ vanishes by (\ref{3c3SD}). Furthermore, using (\ref{3c1}) and (\ref{3c3}) this torsion simplifies to
\begin{eqnarray}
\mathfrak{T}_{ab}{}^{\gamma k} = -\tfrac{2i}{5}\, (\gamma_{ab})_{\beta}{}^{(\gamma} \mathcal D_{\delta}{}^{k} N^{\delta) \beta}
        						     +\tfrac{3i}{5}\,  (\gamma_{ab})_{\beta}{}^{[\gamma} \mathcal D_{\delta}{}^{k} N^{\delta] \beta}
						 = -\tfrac{i}{2}\, (\gamma_{ab})_{\beta}{}^{\delta} \mathcal N_{\delta}{}^{k\, \beta \gamma} ~.
\end{eqnarray}
It is easily verified that this combination is $\gamma$-traceless due to the tracelessness of $\mathcal N$. Additionally, using the constraint relations (\ref{constraint2}) and (\ref{constraint3}), we obtain 
\begin{eqnarray}
\mathfrak{T}_{a\, \beta j} = -\tfrac{7i}{4} \, \mathcal C_{a\, \beta j}~.
\end{eqnarray}

 In order to finish the analysis of the dimension-$\tfrac{3}{2}$ Bianchi identities, it remains to decompose the field strength (\ref{dim3/2F}). Expanding 
\begin{eqnarray}
F_{a\, \gamma k}{}^{ij}= \mathfrak{F}_{a\, \gamma k}{}^{ij} + (\gamma_a)_{\gamma \delta} \, \mathfrak{F}^{\delta}{}_{k}{}^{ij}
             			      + \delta_{k}^{(i} \,\mathfrak{F}_{a\, \gamma}{}^{j)}  
			               + \delta_{k}^{(i} \, (\gamma_a)_{\gamma \delta} \, \mathfrak{F}^{\delta i)} ~,
\end{eqnarray}
we may resolve the field strength into its irreducible components, by projections of the equation\footnote{This expression follows simply from symmetries arguments.}
\begin{eqnarray}
F_{\alpha k\, c\, ij} = F_{\alpha (k\, c\, ij)} - \tfrac{2}{3}\, \varepsilon_{k(i|} F_{\alpha l \, c\, | j)}{}^{l} ~.
\end{eqnarray}
This gives
\begin{align}
 \mathfrak{F}_{a\, \gamma k \, ij} &=\tfrac{1}{2}\, \tau_{a \, \gamma}^{b\, \delta}(5,1) \mathcal D_{\delta (k} C_{b) ij} =0~, \\
 \mathfrak{F}^{\delta}{}_{k\, ij} &=-\tfrac{1}{6}\, \mathcal D_{\gamma(k}C^{\gamma \delta}{}_{ij)} = - \mathcal C^{\delta}{}_{ijk}~, \\
 \mathfrak{F}_{a\, \gamma k} &= \tfrac{1}{3}\, \tau_{a \, \gamma}^{b\, \delta}(5,1) \mathcal D_{\delta}{}^{l} C_{b\, lk}  
                                                      - \tfrac{4}{3}\, (\tilde \gamma_{a})^{\beta \delta}\, \mathcal D_{\beta k} N_{\delta \gamma} ~,\\
 \mathfrak{F}^{\alpha i} &=\tfrac{5}{9}\,\mathcal D_{\beta j} C^{\beta \alpha ij} + \tfrac{2}{3}\, \mathcal D_{\beta}{}^{i} N^{\beta\alpha}     
 				      = -5\, \mathcal C^{\alpha i} + \tfrac{5}{3} \, \mathcal N^{\alpha i}  ~.
 \end{align}
where the term $\mathfrak{F}_{a\, \gamma k}$ can be simplified using (\ref{3c2}) and (\ref{constraint3.1}) to
 \begin{eqnarray}
 \mathfrak{F}_{a\, \gamma k}= 3\, \mathcal C_{a\, \gamma k}
  						 - \tfrac{4}{3} (\tilde \gamma_{a})^{\delta \beta} \mathcal D_{\delta k} N_{\beta \gamma} 
				             = 6\, \mathcal C_{a\, \gamma k}~.
 \end{eqnarray}
At this point the only irreducible tensors which have not been simplified are $\mathfrak{T}^{\alpha i}$ and $\mathfrak{F}^{\alpha i}$. These combinations involve constraints on the self-dual part of the superfield $N$ which, as we will see in chapter \ref{ConformalStructure}, is covariant under conformal transformation (and therefore, the superspace version of the Weyl tensor is constructed from it).

\section{Dimension-2 Bianchi identities}
In this section we study the dimension-2 Bianchi identities. As indicated in table (\ref{Table2.1}), there are four pieces with this dimension: The parts proportional to the Lorentz and SU(2) generators within the $[ssv\}$-indentity, the part proportional to the spinorial covariant derivative inside the $[svv\}$-identity, and the piece proportional to the vector derivative appearing in the $[vvv\}$-identity. The latter, is identically fulfilled, giving rise to what is known as  `` Second Bianchi Identity" for the Riemann tensor
\begin{equation}
R_{a [bcd]} =0~.
\end{equation}
Let us proceed with the study of the first three aforementioned identities. The part proportional to the Lorentz generator $M_{ab}$ within the $[ssv\}$-indentity (\ref{ssv}) is given by
\begin{align}
0&= i\:\varepsilon_{ij} (\gamma^a)_{\alpha\beta} \left[ R_{ca}{}^{bd} + 4 \mathcal D_{c} N_{a}{}^{bd}\right] + 2i\:(\gamma^{abd})_{\alpha\beta} \mathcal D_c C_{aij} 
   +  \mathcal D_{(\underline \alpha} R_{\underline \beta) ) c}{}^{bd}  \nonumber \\
 &+ 4i\: T_{(\underline \alpha \,c}{}^{\gamma k} \left[ (\gamma^{abd})_{\beta) \gamma} C_{aj)k} + 2 \varepsilon_{j)k} (\gamma_{a})_{\beta)\gamma} N^{abd}  \right] ~.
 \label{ssvM}
\end{align}
The above expression is symmetric in composite indices $_{(\underline{\alpha} \underline{\beta})}$. Such a symmetry can be implemented through simultaneous symmetry or antisymmetry of both, spin and isopin indices. Let us first  analize the double antisymmetric case. We can isolate the Riemann tensor multiplying (\ref{ssvM}) by $\tfrac{i}{8}\varepsilon^{ij} (\tilde \gamma_e)^{\alpha\beta}$. This gives
\begin{align}
R_{ce}{}^{bd} =&- 4\: \mathcal D_c N_{e}{}^{bd} 
            - \tfrac{i}{8} (\tilde \gamma_e)^{\alpha\beta}\: \mathcal  D_{\alpha i} R_{\beta}{}^{i}{}_{c}{}^{bd} \cr
  &- \tfrac{1}{2} (\tilde \gamma_e)^{\alpha\beta}\: T_{\alpha}{}^{j}{}_{ c}{}^{\gamma k} \left[ (\gamma^{abd})_{\beta\gamma} C_{ajk} 
  + 2 \varepsilon_{jk} (\gamma^{a})_{\beta\gamma} N_{a}{}^{bd}  \right]    ~,
  \label{riemann1}
  \end{align} 
and plugging the curvature and torsion  back in (\ref{riemann1}), we obtain a first expression for the Riemann tensor
\begin{align}
R_{ce}{}^{bd} &= - 4\: \mathcal D_c N_{e}{}^{bd} + \tfrac{1}{8}(\gamma_{ce})_{\delta}{}^{\alpha}  \mathcal D_{\alpha i} T^{bd \delta i} 
 - \tfrac{1}{8} \eta_{ce} \mathcal D_{\alpha i} T^{bd \alpha i} - \tfrac{1}{4}  \mathcal D_{\alpha i} (\gamma^{[d}{}_{e})_{\delta}{}^{\alpha} T_{c}{}^{b] \delta i}  \nonumber \\
  &+ \tfrac{1}{4}  \mathcal D_{\alpha i}\, \delta_{e}^{[d} \, T_{c}{}^{b] \alpha i} - \tfrac{1}{2} \text{tr} (\gamma^{abd} \tilde \gamma _{fce} ) \, C_{ajk} C^{fjk} 
   + 16 \, N_{ace} N^{abd} ~.
\label{riemann2}
\end{align}
 Symmetries of the curvature tensor (\ref{riemann2}) should be fulfilled. On the one hand, clearly $R_{ce}{}^{(bd)} = 0 $ identically. On the other hand, demanding $R_{(ce)}{}^{bd}=0$ we get 
\begin{equation}
 \mathcal D_{(c} N_{e)}{}^{bd} = - \tfrac{1}{32} \eta_{ce} \mathcal D_{\alpha i} T^{bd \alpha i} 
 						      - \tfrac{1}{16}  \mathcal D_{\alpha i} (\gamma^{[d}{}_{(e})_{\delta}{}^{\alpha} T_{c)}{}^{b] \delta i} 
						      +\tfrac{1}{16}  \mathcal D_{\alpha i}\, \delta_{(e}^{[d} \, T_{c)}{}^{b] \alpha i} ~,
\label{symriemann}
\end{equation}
equation which can be contracted with the metric tensor $\eta^{ce}$ to obtain the divergence of the superfield $N$
\begin{equation}
\mathcal D_{c} N^{cbd} =
                \tfrac{1}{16}\: \mathcal D_{\alpha i} (\gamma^{c[d})_{\delta}{}^{\alpha} T_{c}{}^{b] \delta i}  - \tfrac{1}{4}\: \mathcal D_{\alpha i} T^{bd \alpha i} ~.
\label{DNeq1}                
\end{equation}
From the Riemann tensor (\ref{riemann2}), we can obtain the Ricci tensor 
\begin{equation}
R_{cb} =  \tfrac{3}{4} \: \mathcal D_{\alpha i} T_{c\,b}{}^{\alpha i} - \tfrac{1}{4} \: \mathcal D_{\alpha i} \: (\gamma_{d(c})_{\delta}{}^{\alpha}  \:  T_{b)}{}^{d \:\delta i} 
             + 8 \: C_{(b}{}^{jk}  C_{c)jk}   -  8\: \eta_{cb} \: C^{ajk} C_{ajk}  
                                                            +  16\, N_{ad(b} N_{c)}{}^{ad} ~.
\label{ricci1}
\end{equation}
Here, we note that requiring the symmetry of the Ricci tensor $R_{[ab]}=0$ is equivalent to $\mathcal D_{\alpha i} T_{ab}{}^{\alpha i}=0$. This reduce the Riemann and the Ricci tensors to
\begin{align}
\label{riemann3}
R_{ce}{}^{bd} = &- 4\: \mathcal D_{[c} N_{e]}{}^{bd} + \tfrac{1}{8}(\gamma_{ce})_{\delta}{}^{\alpha}  \mathcal D_{\alpha i} T^{bd \delta i} 
 - \tfrac{1}{4}  \mathcal D_{\alpha i} (\gamma^{[b}{}_{[c})_{\delta}{}^{\alpha} T_{e]}{}^{d] \delta i}   \cr
 &- \tfrac{1}{2} \text{tr} (\gamma^{abd} \tilde \gamma _{fce} ) \, C_{ajk} C^{fjk}  
 + 16 \, N_{ace} N^{abd}    ~, 
\end{align}
and
\begin{equation}
\label{ricci3}
R_{cb} = - \tfrac{1}{4} \: \mathcal D_{\alpha i} \: (\gamma_{d(c})_{\delta}{}^{\alpha}  \:  T_{b)}{}^{d \:\delta i} 
             + 8 \: C_{(b}{}^{jk}  C_{c)jk}  - 8\: \eta_{cb} \: C^{ajk} C_{ajk}  +  16\, N_{ad(b} N_{c)}{}^{ad}  ~.
\end{equation}
The Ricci scalar arises directly from (\ref{ricci1})
\begin{eqnarray}
R=\tfrac{1}{4}\:  (\gamma_{ab})_{\delta}{}^{\alpha}\: \mathcal D_{\alpha i}  T^{ab \delta i} - 40 \: C_{aij}C^{aij} + 16\: N_{abc} N^{abc} ~. 
 \label{scalar}
\end{eqnarray}
The above curvature quantities depends on a combination which involves the (spin) derivative of the dimension-$\tfrac{3}{2}$ 
torsion, $\Delta_{ab}{}^{cd}:=(\gamma_{ab})_{\delta}{}^{\alpha}  \mathcal D_{\alpha i} T^{cd \delta i}$, and further symmetries and contractions of it. It is, in general, not direct to express such combination in terms of irreducible pieces. For this reason, as we will see, it will be simpler to compute the Riemann tensor from the $[svv\}$-identity. Nevertheless, it is possible at this moment to write down 
$\Delta_{a(bc)}{}^{d}$ and $\Delta_{ab}{}^{ab}$ in terms of irreducible parts, and therefore the Ricci tensor (\ref{ricci3}), together with the curvature scalar (\ref{scalar}) are given by
\begin{align}
\label{ricci4}
R_{ab} =&\tfrac{i}{8}  \eta_{ab} \left[ 10\: \mathcal D_{\alpha i} \mathcal C^{\alpha i} - \tfrac{5}{3}\: \mathcal D_{\alpha i} \mathcal N^{\alpha i} +64i \: C^{dij} C_{dij}  \right]
                                 + 8 \: C_{a}{}^{ij}  C_{bij}   
             +  16\, N^{cd}{}_{a} N_{bcd}~,  \\
 \label{scalar2}
R=&\tfrac{15i}{2}\: \mathcal D_{\alpha i} \mathcal C^{\alpha i}  - 40 \: C_{aij}C^{aij}
           -\tfrac{5i}{4}\: \mathcal D_{\alpha i} \mathcal N^{\alpha i} + 16\: N_{abc} N^{abc}~.
\end{align}
This completes the analysis of the double antisymmetric part of (\ref{ssvM}). From the double symmetric side, we can isolate the 
$\mathcal D C$ term by contracting (\ref{ssvM}) with the 3-form $(\tilde \gamma^{c}{}_{bd}) ^{\alpha \beta}$. This gives
\begin{align}
0  =  2i\, \text{tr} (\tilde \gamma^{c}{}_{bd} \gamma^{abd} )   \mathcal D_{c} C_{aij} 
  &+ (\tilde \gamma^{c}{}_{bd}) ^{\alpha \beta} \mathcal D_{\alpha (i} R_{\beta j) c}{}^{bd}   \\
   &+4i \; (\tilde \gamma^{c}{}_{bd}) ^{\alpha \beta} T_{\alpha (i c}{}^{\gamma k }
           \left[ (\gamma^{abd})_{\beta\gamma} C_{aj)k} 
            + 2 \varepsilon_{j)k} (\gamma^{a})_{\beta\gamma} N_{a}{}^{bd}  \right]   ~,    \nonumber        
\label{ssvM2}
\end{align}
which gives the divergence of the $C$-field
\begin{equation}
\mathcal D_{a} C^{aij} =  \tfrac{i}{12} \: (\tilde \gamma_a)^{\alpha\beta} \mathcal D_{\alpha}{}^{(i} \mathcal D_{\beta k} C^{aj)k}
                                        =  \tfrac{3i}{4} \mathcal D_{\alpha}{}^{(i} \mathcal C^{\alpha j)} ~.
\label{DCeq1}
\end{equation}
For the sake of completeness,  we can also compute the divergence of the $N$-field. From the symmetries of the Ricci tensor, we argued that  $\mathcal D_{\alpha i} T_{ab}{}^{\alpha i}=0$. Combining this constraint with Eq. (\ref{DNeq1}) one obtain
\begin{equation}
 \mathcal D^{c} N_{abc}= 
      \tfrac{i}{8}  (\gamma_{ab})_{\beta}{}^{\alpha} \left[ \mathcal D_{\alpha i} \mathcal C^{\beta i}
       +\tfrac{5}{12}\, \mathcal D_{\alpha i} \mathcal N^{\beta i} \right]~.
\label{DN}      
\end{equation}
This completes the analysis of the piece proportional to the Lorentz generator $M$ within the $[ssv\}$-identity.\\
Next, we focus on the part proportional to the SU(2) generator, $J_{ij}$,  in (\ref{ssv}). This is
\begin{eqnarray}
0&=&2i\:\varepsilon_{ij}(\gamma^a)_{\alpha\beta}\left[ F_{ca}{}^{lm}-3 \mathcal D_{c} C_{a}{}^{lm}\right]
             -16i\:\delta^{(l}_{i}\delta^{m)}_{j}\mathcal D_{c}N_{\alpha\beta}
             +2\:\mathcal D_{(\alpha (i} F_{\beta) j) c}{}^{lm}   \nonumber \\
&+&4i\: T_{c (\alpha(i }{}^{\gamma k} \left[3\:\varepsilon_{j)k} (\gamma_a)_{\beta)\gamma} C^{alm} + 8\:\delta^{(l}_{j)}\delta^{m)}_{k} N_{\beta)\gamma}\right] ~.
\label{ssvJ}
\end{eqnarray}
Naturally, in the same way that the part proportional to the Lorentz generation, equation (\ref{ssvJ}) exhibits the symmetry 
$_{(\underline{\alpha} \underline{\beta})}$, which may be realized through a double symmetry or antisymmetry of spin and isospin indices.  In order to obtain the SU(2)- field strength, we proceed to focus on the double antisymmetry of (\ref{ssvJ}). Contracting with $\tfrac{i}{16}\varepsilon^{ij}(\tilde \gamma)^{\alpha\beta}$ the second term vanish due the $_{(ij)}$-symmetry and we can isolate $F_{ab}{}^{ij}$. The resulting expression is not easily expressed explicitly in terms of  fundamental superfields. Therefore, as well as for the Riemann tensor, we will see that it will be more manageable to compute the SU(2) field strength from the $[svv\}$-identity. Nevertheless, at this point, the antisymmetry $F_{(ab)}{}^{ij}=0$ is required, 
obtaining the divergence of the $C$-field
\begin{equation}
 \mathcal D_{a} C^{alm} 
         = \tfrac{i}{4} \: (\tilde \gamma_a)^{\alpha \beta} \mathcal D_{\alpha}{}^{(l} \mathcal D_{\beta j}  C^{am) j}    
            =  \tfrac{9i}{4} \mathcal D_{\alpha}{}^{(i} \mathcal C^{\alpha j)} ~.
 \label{DCeq2}
 \end{equation}
Then, comparing (\ref{DCeq1}) and (\ref{DCeq2}) we see that the $C$ superfield is divergence-free, that is $ \mathcal D_{a} C^{aij}=0=\mathcal D_{\alpha}{}^{(i} \mathcal C^{\alpha j)} $. \\
Finally, the double symmetric combination of spin and isospin indices in (\ref{ssvJ}) can be considered by multiplying by the three form $(\tilde \gamma^{cde})^{\alpha \beta}$ and contracting isospin indices. This gives
\begin{equation}
0=\mathcal D^c N^{(-)}_{abc}+8 \, N^{(+)cd}{}_{[a} N^{(-)}_{b]cd} 
          +\tfrac{11i}{32}\, (\gamma_{ab})_{\beta}{}^{\alpha} \mathcal D_{\alpha i} \mathcal C^{\beta i}    
                - \tfrac{5i}{96}\, (\gamma_{ab})_{\beta}{}^{\alpha} \mathcal D_{\alpha i} \mathcal N^{\beta i}   ~.
\label{DNeq5}                
\end{equation}
This equation can be combined with (\ref{DN}) in order to obtain some expressions for the divergence of the superfield $N$. 
\begin{align}
(\gamma_{ab})_{\beta}{}^{\alpha} \mathcal D_{\alpha i} \mathcal C^{\beta i} &=  -\tfrac{16 i}{3} \mathcal D^c \tilde N _{abc}~,  \\
(\gamma_{ab})_{\beta}{}^{\alpha} \mathcal D_{\alpha i} \mathcal N^{\beta i} &=  -\tfrac{96 i}{5} 
            							\left[  \mathcal D^c N _{abc} - \tfrac{2}{3}\, \mathcal D^c \tilde N _{abc}     \right]  ~,  \\
\label{EOM?}							
16\, N^{(+)cd}{}_{[a} N^{(-)}_{b]cd} &= -3\, \mathcal D^c N^{(+)}_{abc} + 5\, \mathcal D^c N^{(-)}_{abc}	~,		
\end {align}
where, $\tilde{N}$ denotes the 3-form dual to $N$. This concludes the study of the dimension-2 part of the Bianchi identity (\ref{ssv}).\\
Finally, the last part to be considered in the dimension-2 analysis  is the part proportional to the spinorial derivative arising from the $[svv\}$-identity (\ref{svv}). This is given by
\begin{align}
0&=\mathcal D_{\alpha i} T_{ab}{}^{\beta j} + \tfrac{1}{4}\, \delta_{i}^{j} \, (\gamma_{cd})_{\alpha}{}^{\beta}\, R_{ab}{}^{cd} 
  + \delta_{\alpha}^{\beta} \, F_{ab \, i}{}^{j}  + 2\, \mathcal D_{[a} T_{b] \alpha i}{}^{\beta j}  \nonumber \\
  &- 2\, (\gamma_{c[a})_{\gamma}{}^{\beta} T_{\, b] \alpha i}{}^{\gamma}{}_{k} \, C^{c\, jk} 
  -2\, (\gamma^{cd})_{\gamma}{}^{\beta} T_{\alpha i [a}{}^{\gamma j} \, N_{b]cd} ~.
\label{svv1}
\end{align}
From here, the Riemann tensor  and the SU(2) field strength will be computed. 
\paragraph{The Riemann tensor} is contained in the second term of (\ref{svv1}). This may be isolated by multiplying the whole expression by 
$(\gamma^{ef})_{\beta}{}^{\alpha}$ and taking the trace $i=j$. This yields \footnote{Note that it is not possible having a term like 
$\varepsilon^{cdmnpq}\, N_{amn}N_{bpq}$ within the Riemann tensor, because such a term is symmetric in $(ab)$. This argument also makes clear why in 6D, necessarily $\tilde{N}_{abc}N^{abc}=0$.} 
\begin{align}
R_{ab}{}^{cd} =& -\tfrac{i}{8} (\gamma^{cd})_{\beta}{}^{\alpha} (\gamma_{ab})_{\gamma}{}^{\delta} 
                             \mathcal D_{\alpha i} \mathcal D_{\delta}{}^{i} N^{\beta \gamma}
             - \varepsilon_{ab}{}^{cdmn} \mathcal D^p\left[ N_{mnp} - \tfrac{5}{2} \tilde{N}_{mnp}   \right]  
  + 8\,\mathcal D_p \delta_{[a}^{[c} N_{b]}{}^{d] p}     \cr
  &- 6\, \mathcal D_p \delta_{[a}^{[c} \tilde{N}_{b]}{}^{d] p}  
    +\tfrac{i}{2} \, \delta_{[a}^{[c} \delta_{b]}^{d]} \mathcal D_{\alpha i} \mathcal C^{\alpha i}
    -\tfrac{5i}{24} \, \delta_{[a}^{[c} \delta_{b]}^{d]} \mathcal D_{\alpha i} \mathcal N^{\alpha i}
   + 8\, \mathcal D_{[a} N_{b]}{}^{cd} \cr
    &-32\, N_{e[a}{}^{[c} N^{d]}{}_{b]}{}^{e}  
    + 8\, \delta_{[a}^{[c} C_{b] ij}C^{d] ij} 
              -4\, \delta_{[a}^{[c} \delta_{b]}^{d]} \, C_{a ij} C^{a ij} ~,
\label{Riem1}                         
\end{align}
so that the only reducible term is the first one. This can be computed by demanding the exchange symmetry $R_{ab\,cd}=R_{cd\,ab}$. From such symmetry, it follows that
\begin{align}
(\gamma^{cd})_{\beta}{}^{\alpha} (\gamma_{ab})_{\gamma}{}^{\delta} 
                            \{ \mathcal D_{\alpha i} ,  \mathcal D_{\delta}{}^{i} \} \,  N^{\beta \gamma}  =
  &-128i\,\mathcal D_p \delta_{[a}^{[c} N_{b]}{}^{d] p}    +96i\, \mathcal D_p \delta_{[a}^{[c} \tilde{N}_{b]}{}^{d] p}  
 - 64i\, \mathcal D_{[a} N_{b]}{}^{cd} \cr
 &+64i\, \mathcal D^{[c} N^{d]}{}_{ab} ~. 
\label{ggDDN}                         
\end{align}
With (\ref{ggDDN}) in hand, it is simple to compute the reducible term in the Riemann tensor (\ref{Riem1})
\begin{align}
 (\gamma^{cd})_{\beta}{}^{\alpha} (\gamma_{ab})_{\gamma}{}^{\delta} 
                             \mathcal D_{\alpha i} \mathcal D_{\delta}{}^{i} N^{\beta \gamma} =&
-(\gamma^{cd})_{\beta}{}^{\alpha} (\gamma_{ab})_{\gamma}{}^{\delta} \mathbf{N}_{\alpha \delta}{}^{\beta\gamma}
-64i\,\mathcal D_p \delta_{[a}^{[c} N_{b]}{}^{d] p}    + 48i \, \mathcal D_p \delta_{[a}^{[c} \tilde{N}_{b]}{}^{d] p}  \cr
  &- 32i\, \mathcal D_{[a} N_{b]}{}^{cd} + 32i\, \mathcal D^{[c} N^{d]}{}_{ab}
   +8i\,\varepsilon_{ab}{}^{cdmn} \mathcal D^p\left[ N_{mnp} - \tfrac{2}{3} \tilde{N}_{mnp}   \right]  \cr
 &-\tfrac{8i}{3}\, \varepsilon_{ab}{}^{cdmn} \left[\mathcal D^p N^{(+)}_{mnp} - 8\, N^{(-)pq}{}_{m}N^{(+)}_{npq}  \right]    
 - \delta_{[a}^{[c} \delta_{b]}^{d]} \, \mathcal D_{\alpha i} \mathcal N^{\alpha i}~,   \nonumber\\
 \label{1stermR}
\end{align}
where $\mathbf{N}_{\alpha \beta}{}^{\gamma\delta}$ stands for the Weyl tensor, defined as
\begin{equation}
\mathbf{N}_{\alpha \beta}{}^{\gamma\delta} = \mathcal D_{(\alpha}{}^{i} \mathcal N_{\beta) i}{}^{\gamma \delta} 
                               - \tfrac{1}{3} \, \delta_{(\alpha}^{(\gamma} \mathcal D_{| \sigma |}{}^{i} \mathcal N_{\beta) i}{}^{\delta) \sigma}~.
\end{equation}
Therefore, replacing (\ref{1stermR}) in (\ref{Riem1}), the Riemann tensor for the supergeometry is obtained
\begin{align}
R_{ab}{}^{cd} &= 
\tfrac{i}{8} (\gamma^{cd})_{\beta}{}^{\alpha} (\gamma_{ab})_{\gamma}{}^{\delta} \mathbf{N}_{\alpha \delta}{}^{\beta\gamma}
+2\, \varepsilon_{ab}{}^{cdmn} \mathcal D^p \left[ N^{(+)}_{mnp} - \tfrac{4}{3}\, N^{(-)}_{mnp}\right] \nonumber\\
&+ 4\, \mathcal D_{[a} N_{b]}{}^{cd} + 4\, \mathcal D^{[c} N^{d]}{}_{ab} 
-32\, N_{e[a}{}^{[c} N^{d]}{}_{b]}{}^{e} + 8\, \delta_{[a}^{[c} C_{b] ij}C^{d] ij} \nonumber\\
&+ \tfrac{i}{2} \, \delta_{[a}^{[c} \delta_{b]}^{d]} \left[ \mathcal D_{\alpha i} \mathcal C^{\alpha i} + 8i\, C_{n ij} C^{n ij}
 -\tfrac{1}{6}\, \mathcal D_{\alpha i} \mathcal N^{\alpha i}   \right] ~, 
 \label{RiemFinal}
\end{align}
where we have used (\ref{EOM?}) in order to write $N^{(-)} N^{(+)}$ in terms of derivatives of the selfdual and antiselfdual part of $N$. As a consistency check, straightforward calculation shows that further contraction of the Riemann tensor (\ref{RiemFinal}) give rise to the Ricci tensor (\ref{ricci4}) and Ricci scalar (\ref{scalar2}).

\paragraph{The SU(2) field strength} can be extracted from (\ref{svv1}) by tracing $\alpha=\beta$ and rearranging isospin indices
\begin{equation}
F_{ab}{}^{ij} = -\tfrac{1}{4}\, \mathcal D_{\alpha}{}^{i}\, T_{ab}{}^{\alpha j} 
                             +\tfrac{1}{2}\, (\gamma_{c[a})_{\beta}{}^{\alpha} T_{b] \alpha }{}^{i \,\beta}{}_{k} \, C^{c \,jk}
                             +\tfrac{1}{2}\, (\gamma^{cd})_{\beta}{}^{\alpha} T_{\alpha}{}^{i}{}_{ [a}{}^{\beta j} \, N_{b]cd}~.
\end{equation}
Demanding $F_{ab}{}^{[ij]}=0$, we get $\mathcal D_{\alpha i} T_{ab}{}^{\alpha i}=0$, in agreement with previous analysis. It also follows, from the symmetric piece in isospin indices, that
\begin{eqnarray}
F_{ab}{}^{ij} = -\tfrac{1}{4}\, \mathcal D_{\alpha}{}^{(i}\, T_{ab}{}^{\alpha j)} - 2\, C_{[a}{}^{k(i} C_{b]}{}^{j)}{}_{k}
                       + 8\, N_{abc}C^{c\, ij} ~,
\label{SU2FS}                       
\end{eqnarray}
where the first term is not irreducible. Taking the derivative of the dimension $\tfrac{3}{2}$-torsion, this term can expressed as follows 
\begin{align}
\label{DT(ij)}
\mathcal D_{\alpha}{}^{(i}\, T_{ab}{}^{\alpha j)} &= 
- \tfrac{5i}{3}\, \mathbf{N}_{ab}{}^{ij} + \tfrac{11i}{72} \, \mathbf C_{ab}{}^{ij} -\tfrac{10i}{9} \, \mathbf{\tilde C}_{ab}{}^{ij}  
   - \tfrac{40}{9} \, \mathcal D_{[a} C_{b]}{}^{ij} - \tfrac{416}{9}\, C_{[a}{}^{k(i} C_{b]}{}^{j)}{}_{k} \nonumber \\
   &  +\tfrac{272}{9}\,N^{(+)}_{abd} C^{dij} - \tfrac{512}{9}\,N^{(-)}_{abd} C^{dij}~,
   \end{align}
where we have defined irreducible superfields
\begin{align}
\mathbf{N}_{ab}{}^{ij} &:= \mathcal D^{(i} \tilde \gamma_{ab} \,\mathcal N^{j)}   ~, \\
\mathbf{C}_{ab}{}^{ij}&:=\mathcal D^{k} \tilde \gamma_{abc} \mathcal D_{k} C^{c}{}_{ij}  ~,   \\
\mathbf{\tilde C}_{ab}{}^{ij}&:= \mathcal D_{[a}{}^{k(i} C_{b] k}{}^{j)} 
				            = \tfrac{1}{4}\, [\mathcal D_{\alpha}{}^{k}, \mathcal D_{\beta}{}^{(i}] (\tilde \gamma_{[a}) C_{b]k}{}^{j)}  ~.
\end{align}
Replacing (\ref{DT(ij)}) into (\ref{SU2FS}), we conclude that the SU(2) field strength of the supergeometry will be given by
\begin{align}
\label{FFinal}
F_{ab}{}^{ij} = \tfrac{5i}{12}\, \mathbf{N}_{ab}{}^{ij} - \tfrac{11i}{288} \, \mathbf C_{ab}{}^{ij} +\tfrac{5i}{18} \, \mathbf{\tilde C}_{ab}{}^{ij}  
   &+ \tfrac{10}{9} \, \mathcal D_{[a} C_{b]}{}^{ij} + \tfrac{86}{9}\, C_{[a}{}^{k(i} C_{b]}{}^{j)}{}_{k} \nonumber \\
   &+ \tfrac{4}{9}\,N^{(+)}_{abd} C^{dij} + \tfrac{200}{9}\,N^{(-)}_{abd} C^{dij} ~.
\end{align}

This result concludes that analysis of the Bianchi identities. 
Summarizing, we have computed completely the geometrical information necessary for the description of simple six-dimensional superspace supergravity. Specifically, we have fixed the dimension-1 and -$\tfrac{3}{2}$ (anti)commutators defining the derivative superalgebra, we have expressed the dimension-$\tfrac{3}{2}$ curvature, field strength and torsion in terms of irreducible parts, and we have computed all the relevant curvature quantities which characterize the supergeometry. A summary containing the most relevant results of this chapter can be found in appendix (\ref{appendixB}).  It is important to point out that we have studied the superspace from an \emph{off-shell} point of view, in the sense that we have isolated its \emph{geometry}  from
 the dynamics of the supergravity fields entering in the superalgebra.

\chapter{Conformal structure}
\label{ConformalStructure}
The description of matter-coupled supergravity theories turns out to be rather complicated. In this respect, superconformal methods represent a simpler approach to the study of such matter-coupled systems. These methods exploit the fact that, among the spacetime symmetries, conformal symmetry is the maximal symmetry of a non-trivial field theory \cite{ColemanMandula}. The underlying idea is to formulate a gauge theory of the superconformal algebra (the supersymmetric extension of the conformal algebra). Such theory contain extra fields which are then eliminated by imposing curvature constraints or by gauge fixing the extra symmetries.  The result is a  gauge theory of the Poincar\'e supersymmetry algebra where the initial extra symmetries are not visible.

In this chapter, we study the conformal structure of the superspace geometry described in chapter (\ref{Supergeometry}). We do so, by following a different route. Instead of considering the superconformal group as the structure group of the theory, we \emph{impose the conformal invariance of the conventional constraints} (\ref{cc1})-(\ref{cc3})\footnote{Superconformal methods to study conformal $4D$, $\mathcal N=1,2$ superspace were used in \cite{Butter:2009cp}. There, the construction relies on considering the full superconformal group as the structure group of theory. Along this line, one might attempt to construct our 6D superspace by de-gauging a conformal supergeometry.}. In particular, we will fix the super-Weyl transformation rules that   superfields ($C_{aij}$ and $N_{abc}$) and covariant derivatives ($\mathcal D_{\alpha i}$ and $\mathcal D_a$) must obey in order to preserve such set of constraints. 

\section{Super-Weyl transformations}

The super-Weyl (sW) trasformations act on the spinorial covariant derivative as
\begin{eqnarray}
\delta \mathcal D_{\alpha i} = \sigma \mathcal D_{\alpha i} + a \, (\mathcal D_{\beta j} \sigma) M_{\alpha}{}^{\beta}   
 										           + b \, (\mathcal D_{\alpha}{}^{j} \sigma) J_{ij} ~,
\label{sWt}										           
\end{eqnarray}
where $\sigma=\sigma(z)$ is, \emph{a priori}, an arbitrary scalar superfield and $a$, $b$ some coefficients that can be determined by requiring the preservation of the the conventional constraints under the above transformation (\ref{sWt}). Let us consider the transformation of the dimension-1 commutator
\begin{align}
\delta \{\mathcal D_{\alpha i} , \mathcal D_{\beta j} \} &= 2\, \{\delta \mathcal D_{(\underline \alpha} \,  , \mathcal D_{\underline \beta)} \}   \nonumber \\
                   &= 2\sigma \{\mathcal D_{\alpha i} , \mathcal D_{\beta j} \}  + b\, \varepsilon_{ij} (\mathcal D_{[\alpha}{}^{k} \sigma) \mathcal D_{\beta] k}
                     + \left(\tfrac{a}{2}+2\right)\, (\mathcal D_{(\underline \alpha}\sigma) \mathcal D_{\underline \beta)}  \nonumber \\
            &- \left(a + \tfrac{b}{2}\right)\left[ (\mathcal D_{\alpha j} \sigma) \mathcal D_{\beta i}+(\mathcal D_{\beta i} \sigma) \mathcal D_{\alpha j}     \right]   \nonumber \\
                   &+ \left[ a\,(\mathcal D_{\beta j}\mathcal D_{\gamma i}\sigma ) M_{\alpha}{}^{\gamma} +b\, (\mathcal D_{\beta j}\mathcal D_{\alpha}{}^{k}\sigma ) J_{ik}
                           + ({}_{\underline \alpha}\leftrightarrow{}_{\underline \beta})  \right]  ~.
\label{sW1}                          
\end{align}
Preservation of the algebra means that the above expression must be equal to
\begin{equation}
\delta \{\mathcal D_{\alpha i} , \mathcal D_{\beta j} \} = 2i \,\varepsilon_{ij} (\gamma^c)_{\alpha \beta} \, \delta \mathcal D_c
                       		+ \tfrac{1}{2}\, \delta R_{\alpha i \beta j}{}^{cd} M_{cd} + \delta F_{\alpha i \beta j}{}^{kl} J_{kl} ~.
\label{sW2}								   
\end{equation}
Therefore, independent pieces in (\ref{sW1}) and (\ref{sW2}) should cancel each other. In particular, matching the terms proportional to the spinorial covariant derivative we get
\begin{align}
0&=-2i \,\varepsilon_{ij} (\gamma^c)_{\alpha \beta} \, \delta \mathcal D_c +  b\, \varepsilon_{ij} (\mathcal D_{[\alpha}{}^{k} \sigma) \mathcal D_{\beta] k}
      + \left(\tfrac{a}{2}+2\right)\, (\mathcal D_{(\underline \alpha}\sigma) \mathcal D_{\underline \beta)} \nonumber \\
      &- \left(a + \tfrac{b}{2}\right)\left[ (\mathcal D_{\alpha j} \sigma) \mathcal D_{\beta i}+(\mathcal D_{\beta i} \sigma) \mathcal D_{\alpha j}     \right] ~.
\label{sW3}      
\end{align}
The previous equation has the symmetry $_{(\underline \alpha \underline \beta)}$, that can be implemented through simultaneous symmetry or antisymmetry of spin and isospin indices. Taking the symmetric part $_{(ij)}$, we obtain the following condition on the coefficient that parametrize the sW transformation
\begin{equation}
2-\tfrac{3a}{2}-b=0 ~.
\label{coef1}
\end{equation}
Taking now the antisymmetric combination, multiplying (\ref{sW3}) by $\varepsilon^{ij}$ we get
\begin{equation}
0=4i \, (\gamma^c)_{\alpha \beta} \, \delta \mathcal D_c +  \left[ \tfrac{5a}{2}+3b+2  \right] (\mathcal D_{[\alpha}{}^{i} \sigma) \mathcal D_{\beta] i} ~.
\label{anti}
\end{equation}
But the transformation of the vector derivative may have a part, besides the homogeneous term, proportional to spinorial covariant derivative, $\delta \mathcal D_c \propto \mathcal D_{\gamma k}$. Such a term should have the structure
\begin{eqnarray}
\delta \mathcal D_c = 2\sigma \mathcal D_c+ i \alpha (\tilde \gamma_c)^{\beta \gamma} (\mathcal D_{\beta}{}^{k}\sigma) \mathcal D_{\gamma k} + \cdots ~,
\label{vector}
\end{eqnarray}
with $\alpha$ some factor to be determined. Then, plugging (\ref{vector}) into (\ref{anti})
\begin{eqnarray}
2+ \tfrac{5a}{2} + 3b=16\alpha ~.
\label{coef2}
\end{eqnarray}
Additionally, in order to elucidate the values of the parameters  $a$ and $b$, we can compute the preservation of the dimension 1/2 conventional constraint, 
$T_{\alpha i \, b}{}^{c}=0$. This is equivalent to setting to zero the part proportional to the vector covariant derivative within the dimension $3/2$ commutator transformation, which is given by 
\begin{align}
\delta [ \mathcal D_{\alpha i}, \mathcal D_b ] |_{\mathcal D_c} &= a\, (\mathcal D_{\beta i}) [M_{\alpha}{}^{\beta}, \mathcal D_b] 
                        							                           + 2 (\mathcal D_{\alpha i} \sigma) \mathcal D_b 
	   +i\alpha (\tilde \gamma_b)^{\beta \gamma} [\mathcal D_{\alpha i}, (\mathcal D_{\beta}{}^{k} \sigma) \mathcal D_{\gamma k}] |_{\mathcal D_c} \nonumber \\
	   &= -\tfrac{a}{2} (\gamma_{bc})_{\alpha}{}^{\beta} (\mathcal D_{\beta i}\sigma) \mathcal D^c + 2(\mathcal D_{\alpha i}\sigma)\mathcal D_b
	    + 2\alpha (\gamma_{c}\tilde \gamma_{b})_{\alpha}{}^{\beta} (\mathcal D_{\beta i}\sigma) \mathcal D^c ~.
\label{delta3/2}	    
\end{align}
Here, the $\gamma\tilde \gamma$ product of the last term decompose as the metric tensor (arising from the symmetric part that satisfy the Clifford algebra) and a 2-form (antisymmetric part), and thus the last term in (\ref{delta3/2}) combines to the first two. Then
\begin{equation}
\delta [ \mathcal D_{\alpha i}, \mathcal D_b ] |_{\mathcal D_c} =
            - \left(\tfrac{a}{2} + 2\alpha \right) (\gamma_{bc})_{\alpha}{}^{\beta} (\mathcal D_{\beta i}\sigma) \mathcal D^c
            + 2(1-\alpha) (\mathcal D_{\alpha i}\sigma)\mathcal D_b ~.
\end{equation}
Demanding that the above expression vanish, we get
\begin{equation}
   \alpha = 1 \, \, \, \text{and} \,\, \, a=-4  ~.
\label{coef3}   
\end{equation}
Therefore, the set of equations (\ref{coef1}), (\ref{coef2}) and (\ref{coef3}) is consistent for
\begin{equation}
b=8~.
\end{equation}
At this point, we have got the sW transformation of the spinorial covariant derivative
\begin{eqnarray}
\label{sWDspin}
\delta \mathcal D_{\alpha i} = \sigma \mathcal D_{\alpha i} -4 \, (\mathcal D_{\beta j} \sigma) M_{\alpha}{}^{\beta}   
 										           +8 \, (\mathcal D_{\alpha}{}^{j} \sigma) J_{ij}		~.
\end{eqnarray}
Let us now focus on the transformations rules for the superfields $C_{a\, ij}$ and $N_{abc}$.  For this, we notice that for the values of $a$ and $b$ we just got, the third an fourth term in (\ref{sW1}) vanish. Thus, requiring  preservation of the algebra of (spinorial) covariant derivatives, that is, equating (\ref{sW1}) and (\ref{sW2}), yields 
\begin{align}
\label{preservation}   
0&= 2i \,\varepsilon_{ij} (\gamma^c)_{\alpha \beta} \, \delta \mathcal D_c + \tfrac{1}{2}\, \delta R_{\alpha i \beta j}{}^{cd} M_{cd} 
      + \delta F_{\alpha i \beta j}{}^{kl} J_{kl}  
   - 2\sigma \{\mathcal D_{\alpha i} , \mathcal D_{\beta j} \}   \nonumber \\
   &- b\, \varepsilon_{ij} (\mathcal D_{[\alpha}{}^{k} \sigma) \mathcal D_{\beta] k} 
   - \left[ a\,(\mathcal D_{\beta j}\mathcal D_{\gamma i}\sigma ) M_{\alpha}{}^{\gamma} +b\, (\mathcal D_{\beta j}\mathcal D_{\alpha}{}^{k}\sigma ) J_{ik}
                           + ({}_{\underline \alpha}\leftrightarrow{}_{\underline \beta})  \right] .  ~                
\end{align}
Again, the symmetry of the latter equation can be realized in two different ways. Taking the terms symmetric in both, spin and isospin indices gives linearly independent terms proportional to the Lorentz generator $M$ and the SU(2) generator $J$. The part proportional to $M$ gives rise to
\begin{eqnarray}
0= 2i\, (\gamma^{abc})_{\alpha \beta} \left( \delta C_{aij} - 2\sigma C_{aij}  \right) M_{bc}  
  - \left[ -\tfrac{a}{4}(\gamma^{bc})_{(\alpha}{}^{\gamma} \mathcal D_{\beta)( j} \mathcal D_{\gamma i)} \sigma +   
    ({}_{\underline \alpha}\leftrightarrow{}_{\underline \beta})\right] M_{bc} ~.
\label{CpropM}    
\end{eqnarray}
Within the square bracket the $\mathcal D \mathcal D$ term splits into a commutator and an anticommutator. Since $\sigma$ does not carry any spin or SU(2) charge, it follows that  $\{\mathcal D_{\alpha( i} , \mathcal D_{\beta j)} \} \sigma=0 $, so that only the commutator part remains. Then we have
\begin{eqnarray}
0= 2i\, (\gamma^{abc})_{\alpha \beta} \left( \delta C_{aij} - 2\sigma C_{aij}  \right) M_{bc}  
+ \tfrac{a}{4} (\gamma^{bc})_{(\alpha}{}^{\gamma} [\mathcal D_{\beta) (j}, \mathcal D_{\gamma i)} ] \sigma \, M_{bc} ~.
\label{CpropM2}
\end{eqnarray}
The commutator must be antisymmetric in its spin indices. This allows us to write this term as
\begin{align}
(\gamma^{bc})_{(\alpha}{}^{\gamma} [\mathcal D_{\beta) (j}, \mathcal D_{\gamma i)} ] \sigma &= 
(\gamma^{bc})_{(\alpha}{}^{\gamma} \, \delta_{\beta)}^{[\mu}\delta_{\gamma}^{\nu]}\,
    [\mathcal D_{\mu (i}, \mathcal D_{\nu j)} ] \sigma  \nonumber \\
&=-\tfrac{1}{4}\, (\gamma^{bc}\gamma^{a})_{(\alpha \beta)} (\tilde \gamma_{a})^{\mu\nu} \,   
     [\mathcal D_{\mu (i}, \mathcal D_{\nu j)} ] \sigma  \nonumber \\ 
&=-\tfrac{1}{2}\, (\gamma^{abc})_{\alpha \beta} \, \mathcal D_{(i}\tilde \gamma_{a} \mathcal D_{j)} \sigma      ~.
\label{comm}
\end{align}
Therefore, plugging (\ref{comm}) into (\ref{CpropM2}) we obtain 
\begin{eqnarray}
0= 2i\, (\gamma^{abc})_{\alpha \beta} \left[  \delta C_{aij} - 2\sigma C_{aij} 
            + \tfrac{ia}{16} \, \mathcal D_{(i}\tilde \gamma_{a} \mathcal D_{j)} \sigma \right] M_{bc}  ~.
\end{eqnarray}
Finally, since this piece must vanish independently of the others, we must set to zero the coefficient of $M$ in the above equation, obtaining the transformation rule for the $C$ superfield 
\begin{equation}
\delta C_{aij} = 2\sigma C_{aij}  - \tfrac{ia}{16} \, \mathcal D_{(i}\tilde \gamma_{a} \mathcal D_{j)} \sigma ~.
\label{sWC}
\end{equation}
Having obtained the transformation of $C$, we now focus on the transformation for the superfield $N$. This rule also arises from
 (\ref{preservation}), but this time taking the symmetric part $_{(\alpha \beta)}$  and $_{(i j)}$ proportional to the SU(2) generator, $J$. This piece gives
\begin{equation}
0= -\tfrac{8i}{3} \, (\gamma^{abc})_{\alpha \beta} \, \left( \delta N_{abc} - 2\sigma N_{abc}  \right) \, J_{ij}
- \left[ \,  b\, \mathcal D_{(\beta (j}\mathcal D_{\alpha)}{}^{k}\sigma \, J_{i)Ák}
                           + ({}_{\underline \alpha}\leftrightarrow{}_{\underline \beta})  \right]        ~.        
\end{equation}
In the last term, only the commutator part contributes. Due to symmetries, we have that
$[\mathcal D_{(\alpha i}, \mathcal D_{\beta)}{}^{k} ] = -\tfrac{1}{2} \, \delta_{i}^{k}\, [\mathcal D_{(\alpha}{}^{l}, \mathcal D_{\beta)l}]$. Furthermore, using the Fierz identity (\ref{Fierz3form}) for the $3$-form, we can rewrite 
\begin{align}
[\mathcal D_{(\alpha}{}^{l}, \mathcal D_{\beta)l}] =
\delta_{(\alpha}^{\mu} \delta_{\beta)}^{\nu} \, [\mathcal D_{\mu}{}^{l}, \mathcal D_{\nu\,  l}] 
  &= \tfrac{1}{48} \, (\gamma^{abc})_{\alpha \beta} \, 
  (\tilde \gamma_{abc})^{\mu \nu}  \,  [\mathcal D_{\mu}{}^{l}, \mathcal D_{\nu\,  l}]   \nonumber\\
 &= \tfrac{1}{24} \, (\gamma^{abc})_{\alpha \beta} \, \mathcal D^{k} \tilde \gamma_{abc} \mathcal D_{k} ~.
\end{align}
With this we obtain
\begin{equation}
0=   -\tfrac{8i}{3} \, (\gamma^{abc})_{\alpha \beta} \, \left[ \delta N_{abc} - 2\sigma N_{abc} 
                        + \tfrac{ib}{128}\, \mathcal D^{k} \tilde \gamma_{abc} \mathcal D_{k} \, \sigma \right] J_{ij} ~.
\end{equation}
As argued previously, this term  must vanish. Cancelation of the factor of $J$ yields to the sW-transformation of the superfield 
$N$
\begin{equation}
\delta N_{abc} = 2\sigma N_{abc} 
                        - \tfrac{ib}{128}\, \mathcal D^{k} \tilde \gamma_{abc} \mathcal D_{k} \, \sigma ~.
\label{sWN}                        
\end{equation}
This completes the analysis of the doubly-symmetric part of the equation (\ref{preservation}). Next, we proceed to study the doubly-antisymmetric part of it. Tracing with $\varepsilon^{ij}$ gives
\begin{align}
0= &-4i\, (\gamma^c)_{\alpha \beta} \, \left( \delta \mathcal D_c -2\, \sigma \mathcal D_c  \right)
        + 2b \, \mathcal D_{[\alpha}{}^{k} \sigma \, \mathcal D_{\beta] k}  \nonumber \\
     &-8i\,  (\gamma_a)_{\alpha \beta} \, \left( \delta N^{abc} -2\, \sigma N^{abc} \right) M_{bc} 
   + \tfrac{a}{2} \, (\gamma^{bc})_{[\alpha}{}^{\gamma} \mathcal D_{\beta]}{}^{k} \mathcal D_{\gamma k} \sigma \, M_{bc}\nonumber\\
  &-12i\,  (\gamma^a)_{\alpha \beta} \, \left( \delta C_{a}{}^{ij} -2\, \sigma C_{a}{}^{ij} \right) J_{ij} 
  + 2b\, \mathcal D_{[\alpha}{}^{i} \mathcal D_{\beta]}{}^{j} \sigma \, J_{ij} ~. 
\end{align}  
From here, we can  isolate the term $\delta \mathcal D_{c}$ contracting spin indices. Multiplying this last equation by the $2$-form 
$(\tilde \gamma^{d})^{\alpha \beta}$ we get 
\begin{align}
\delta \mathcal D_a &= 2\sigma \mathcal D_a - \tfrac{ib}{8} (\mathcal D^k \sigma)\tilde \gamma_a \mathcal D_{k}
   + 3\, \left[\delta C_{aij} - 2\sigma C_{aij} 
  - \tfrac{ib}{24} (\mathcal D_{i}\tilde \gamma_{a} \mathcal D_{j}  \sigma)\right] J^{ij}      \nonumber \\   
  &-2 \left[ \delta N_{abc} - 2\sigma N_{abc} 
      - \tfrac{ia}{64}  (\mathcal D^{k}\tilde \gamma_{abc} \mathcal D_{k}  \sigma)
      + \tfrac{ia}{32}  \eta_{ab}(\mathcal D^{k}\tilde \gamma_{c} \mathcal D_{k}  \sigma)
                \right] M^{bc} ~.
\end{align} 
Using our  results about transformation laws of $C$ and $N$ this simplifies further to
\begin{align}
\delta \mathcal D_a = 2\sigma \mathcal D_a - \tfrac{ib}{8} (\mathcal D^k \sigma)\tilde \gamma_a \mathcal D_{k}
&+ \tfrac{i}{64} \left[  (2a + b) \,  (\mathcal D^{k}\tilde \gamma_{abc} \mathcal D_{k}  \sigma)
      -  32i\, a \,  \eta_{ab}(\mathcal D_c \sigma) \right] M^{bc}    \nonumber \\
    &-   \tfrac{i}{8}\, \left( \tfrac{3a}{2} + b \right) \, (\mathcal D_{i}\tilde \gamma_{a} \mathcal D_{j}  \sigma)  J_{ij}    ~,      
 \label{vectorsW}   
\end{align}
where we have also used $(\mathcal D^{k}\tilde \gamma_{a} \mathcal D_{k} )\sigma =  8i \, \mathcal D_a \sigma$. The values of the parameters above imply, on the one hand, that $2a+b=0$ and therefore the factor of the $3$-form in the first line of 
(\ref{vectorsW}) vanish. This means that there is no $(\mathcal D^{k}\tilde \gamma_{abc} \mathcal D_{k}  \sigma)M^{bc}$-term within the sW-transformation rule of the bosonic covariant derivative. On the other hand, $(3a+2b)/2 = 2$ so that the vector covariant derivative will transform as
\begin{equation}
\label{sWDvector}
\delta \mathcal D_a = 2\, \sigma \, \mathcal D_a -i\, (\mathcal D^k \sigma)\, \tilde \gamma_a \, \mathcal D_k
    - 2\, (\mathcal D^b \sigma)\, M_{ab} - \tfrac{i}{4} \, (\mathcal D^{i}\tilde \gamma_{a} \mathcal D{j}  \sigma)  J_{ij}   ~.
\end{equation}
This conclude the analysis of the conformal transformations. As a final comment, note that the Weyl transformation rules of the fields (\ref{sWC}) and (\ref{sWN}) contain inhomegeneous terms. Such terms can be used to gauge away some of the components of these superfields.
\chapter{Field content}
\label{FieldContent}
In this chapter we focus on the study of the field content of the six-dimensional conformal supergeometry presented in the previous chapters. We first  review briefly the construction of the Weyl multiplet of Bergshoeff \emph{et alia} \cite{Bergshoeff:1985mz}, which emerges as a realization of the conformal supersymmetry algebra. We then explore how this multiplet appears in superspace.

\section{The Weyl multiplet}
The Weyl multiplet refers to the set of fields on which the six-dimensional superconformal algebra $\text{Osp}(6,2|1)$ is realized. The generators of this algebra are the usual Poincar\'e plus SU(2) generators, as specified at the beginning
of chapter \ref{Supergeometry}
\begin{equation}
\label{generators1}
M_{ab} ~,  ~~  P_{a} ~,  ~~ J_{ij}
\end{equation}
together with the supersymmetry generators plus the dilatation, special conformal, and special supersymmetry generators 
\begin{eqnarray}
\label{generators2}
Q_{\alpha i}~, ~~D~,~~ K_{a}~, ~~ S_{\alpha i} ~.
\end{eqnarray} 
As pointed out in \cite{Bergshoeff:1985mz}, the superconformal algebra generated by (\ref{generators1}) and (\ref{generators2}) can be realized on the following set of fields
\begin{align}
\label{typeI}
\begin{array}{ccccccccc}
e^a_m ~~~& \psi_m^{\alpha i} ~~~&\Phi_m^{ij} ~~~ &B_m  ~~~ & N^{(-)}_{abc}~~~ &\chi^{\alpha i} ~~~&F\\
14 ~~~& -32~~~ & 15 ~~~	&0~~~&10~~~	&-8 ~~~& 1		
\end{array}
\end{align}
 Here, the first four fields are the gauge fields corresponding to the generators $P_{a}$, $Q_{\alpha i}$, $J_{ij}$ and $D$, respectively;  $e_{m}^{a}$ is the (inverse of the) frame field, $\psi_m^{\alpha i}$ is the gravitino, $\Phi_m^{ij}$ the SU(2) connection and $B_m$ is the dilatation gauge field. The anti-self-dual tensor $N^{(-)}_{abc}$, the spinor $\chi^{\alpha i}$ and the scalar $F$ are matter fields.
The number of the \emph{off shell} degrees of freedom carried by the field is indicated explicitly. For the gauge fields, the counting of these degrees of freedom can be worked out by counting the number of components of each field and then  subtracting the gauge transformations\footnote{More precisely, the number of the independent degrees of freedom in each gauge parameter entering in the gauge transformations.}
\begin{align}
\delta e_{m}^{a} &= \partial_{m} \xi^{a} + \Lambda^{a}{}_{b} \, e_{m}^{b} + \sigma \, e_{m}^{a} ~, \\
\delta \psi_{m}^{\alpha i} &= \partial_{m}\lambda^{\alpha i} + e^{a}_{m} (\tilde \gamma_{a})^{\alpha \beta} \eta_{\beta}^{i}~,  \\
\delta \Phi_{m}^{ij} &= \partial_{m} \alpha^{ij} ~,\\
\delta B_{m} &= e_{m}^{a} b_{a}.
\end{align}
In this way, to the 36 components of $e^{a}_{m}$, we need to subtract the $6+15+1$ components of the gauge parameters $\xi^{a}$, 
$\Lambda^{a}{}_{b}$ and $\sigma$, respectively, resulting in $36-22=14$ off shell degrees of freedom. In the case of the gravitino 
$\psi_{m}^{\alpha i}$, to its $-48$ components (the minus sign denote fermionic components) we need to subtract the $-8-8$ components of the gauge parameters $\lambda^{\alpha i}$ and $\eta_{\beta}^{i}$, for a total of  $-48+16=-32$ off shell degrees of freedom. Next, the counting for the SU(2) gauge field  $\Phi_{m}^{ij}$ is $18$ components minus the $3$  components of the parameter $\alpha^{ij}$, giving 15 off shell degrees of freedom. Finally, the dilaton gauge field $B_{m}$ is pure gauge, since the gauge parameter $b_{a}$ has the same number of components of it (that is, 6).\\
The matter fields, on the other hand,  carry just their component degrees of freedom with the exception of the anti-self-dual tensor 
$N^{(-)}_{abc}$, which carry only a half of the possible 20 carried by a totally antisymmetric tensor $N_{abc}$ (the remaining 10 components are carried by its  self-dual counterpart $N^{(+)}_{abc}$).

\section{The Weyl multiplet in superspace}
We would now like to understand how the Weyl multiplet just described appears in superspace. Firstly, the component gauge fields plus the gravitino are related to the $\theta = 0$ components of the superframe field and superconnections, while the matter fields  are given by 
\begin{eqnarray}
\label{theta0type1}
N^{(-)}_{abc} = N_{abc} |_{\theta = 0} ~, ~~~ \chi^{\alpha i} = \mathcal N^{\alpha i} |_{\theta = 0} ~, ~~~
     F=\mathcal D_{\alpha i}  \mathcal N^{\alpha i} |_{\theta = 0}
\end{eqnarray}
Secondly, the definition of $\mathcal C^{\alpha i}$ and $\mathcal N^{\alpha i}$ imply that their derivatives decompose as 
\begin{align}
\label{DCdec}
\mathcal D_{\alpha i}\mathcal C^{\alpha}{}_{j}&= \tfrac{8i}{3} \mathcal D_{a} C^{a}{}_{ij} 
 										- \tfrac{1}{2} \varepsilon_{ij} \mathcal D_{\alpha k} \mathcal C^{\alpha k}~,    \\
\label{DNdec}										
\mathcal D_{\alpha i}\mathcal N^{\alpha j}&=\tfrac{1}{2} \, \delta_{i}^{j}\,\mathcal D_{\gamma k}\mathcal N^{\gamma k} ~.
\end{align}
It also follows that the supergravity fields obey the relations
\begin{align}
\label{DC=DNNN}
(\gamma_{ab})_{\alpha}{}^{\beta}\, \mathcal D_{\beta i}\mathcal C^{\alpha i} &=
        -\tfrac{32i}{9}\, \left[ \mathcal D^c N^{(-)}_{abc} - 8\, N^{(+)\: cd}{}_{[a} N^{(-)}_{b] cd} \right]~, \\
\label{DN=DNNN}        
(\gamma_{ab})_{\alpha}{}^{\beta}\, \mathcal D_{\beta i}\mathcal N^{\alpha i}&= 
       -\tfrac{32i}{5}\left[ \mathcal D^c N^{(+)}_{abc} + 8\, N^{(+)\: cd}{}_{[a} N^{(-)}_{b] cd} \right]  ~.
\end{align}
The importance of these expressions is that (\ref{DCdec}) and (\ref{DNdec}) imply that there are no auxiliary iso-triplets $D^{ij}$ in the supergravity multiplet, while from (\ref{DC=DNNN}) and (\ref{DN=DNNN}) follow that there is no new singlet 2-form field strength. However, we do have the 2-forms iso-triplets 
\begin{align}
\mathbf{C}_{ab\, ij} &:= \mathcal D^{k} \, \tilde \gamma_{abc} \, \mathcal D_{k} \, C^{c}{}_{ij} ~,  \\
\mathbf{N}_{ab\, ij} &:= \mathcal D_{(i} \, \tilde \gamma_{ab} \, \mathcal N_{j)} ~,
\end{align}
and the isospin components $\mathbf{C}^{ijkl} := \mathcal D_{\alpha}{}^{(i} \mathcal C^{\alpha jkl)}$. Thus, at this stage we are left with the following set of components fields
\begin{align}
\label{AllComp}
\begin{array}{llllllllllllllllllll}
C^a_{ij}~~~& \mathcal C^\alpha_{ijk}~~~ & \mathcal C_{a\, ij}~~~ & \mathcal C^{\alpha i}~~~ & \mathbf C~~~
	&\mathbf C_{ab\, ij}~~~& \mathbf C_{ijkl}\\
N^{\alpha \beta}&& \mathcal N_{\gamma k}{}^{\alpha \beta} & \mathcal N^{\alpha i} & \mathbf N 
	&\mathbf N_{ab\, ij}& \mathbf N_{\gamma\delta}{}^{\alpha \beta} \\
N_{\alpha \beta}
\end{array}
\end{align}
where we have renamed $\mathbf{C}:=\mathcal D_{\alpha i} \mathcal C^{\alpha i}$ and
$\mathbf{N}:=\mathcal D_{\alpha i} \mathcal N^{\alpha i}$,  and the superfield 
$\mathbf{N}_{\gamma \delta}{}^{\alpha \beta} := \mathcal D_{(\alpha}{}^{i}\mathcal N_{\beta) i}{}^{\gamma \delta}$- traces, denotes the Weyl tensor. Of these fields, one can use the various components in $\sigma$ to gauge away $C^a_{ij}$, $\mathcal C^\alpha_{ijk}$, 
$ \mathcal C_{a\, ij}$, $ \mathbf C_{ijkl}$ and $N_{\alpha \beta}$. This leaves
\begin{align}
\begin{array}{llllllllllllllllll}
	& && \mathcal C^{\alpha i}~~~ & \mathbf C~~~
	&\mathbf C_{ab\, ij}&\\
N^{\alpha \beta}&& \mathcal N_{\gamma k}{}^{\alpha \beta} & \mathcal N^{\alpha i} & \mathbf N 
	&\mathbf N_{ab\, ij}& \mathbf N_{\gamma\delta}{}^{\alpha \beta} 
\end{array}
\end{align}
The bottom row contains the correct component content to describe an anti-self-dual tensor, the curl of the gravitino (both the $\gamma$-traceless and $\gamma$-trace parts), the SU(2) field strength, the curvature scalar, and the Weyl tensor.\\

\chapter{Matter couplings}
\label{MatterCouplings}
This chapter is devoted to the study of the possible matter field configurations compatibles with the conformal superspace structure developed in the previous chapters. In the final part, it is also shown that the constraints defining the scalar (hyper) and tensor multiplets imply a Weyl-type and scalar equation of motion for the superfield defining each multiplet.
\section{Abelian vector multiplet}
Let $W^{\beta(s)\, j(f)}:=W^{\beta_1\cdots \beta_s \, j_1\cdots j_f}$ be an arbitrary superfield of Weyl-weight $w$, symmetric in $s$ spin and $f$ isospin indices,\footnote{We assume total symmetry in both kind of indices, spin and isospin.} with $s\geqslant1$ and 
$f \geqslant1$. The Weyl transformation of the spinorial derivative of the field is given by 
\begin{align}
\delta \left( \mathcal D_{\alpha i}  W^{\beta_1 \cdots  \beta_s\,j_1 \cdots j_f}\right)  &=  
              \sigma (1+2w)  \mathcal D_{\alpha i} W^{\beta_1 \cdots  \beta_s \, j_1 \cdots j_f} + 
              (2w+ s) (\mathcal D_{\alpha i} \sigma)  W^{\beta_1 \cdots  \beta_s \, j_1 \cdots j_f}  \nonumber \\                                                                                                                   
  &- 4  (\mathcal D_{\gamma i} \sigma)  \sum_{q=1}^{s}  \delta ^{\beta_q}_{\alpha}W^{\beta_1 \cdots \beta_{q-1} \gamma \beta_{q+1} \cdots \beta_s \,j_1 \cdots j_f}   \nonumber  \\
   &-8  (\mathcal D_{\alpha}{}^{l} \sigma)  \sum_{q=1}^{f}   \delta ^{j_q}_{(i} W^{\beta_1 \cdots \beta_s \, j_1\cdots  j_{q-1}} {}_{l)}{}^ {j_{q+1} \cdots j_f} ~,
\label{Wupper}
\end{align}
where we have used the following commutators
\begin{align}
[ M_{\alpha} {} ^{\gamma} ,  W^{\beta_1\cdots \beta_s} ] &= -\tfrac{s}{4} \: \delta_{\alpha}{}^{\gamma} W^{\beta_1\cdots \beta_s} 
                    + \sum_q \delta_{\alpha}{}^{\beta_q} W^{\beta_1\cdots  \beta_{q-1} \gamma \beta_{q+1} \cdots \beta_s} ~,
\label{MW} \\
{}[ J_{ij}, W^{k_1\cdots k_f} ]  &=  - \sum _q \delta_{(i}{}^{k_q} W^{k_1\cdots  k_{q-1} }{} _{j)}{}^ { k_{q+1} \cdots k_f}~.
\label{JW}
\end{align}
Contracting $\alpha=\beta_1$ and $i=j_1$, transformation (\ref{Wupper}) becomes (we only need to be careful with the first term of each sum, and split the last isospin sum into its symmetric parts) 
\begin{align}
\delta \left( \mathcal D_{\alpha i}  W^{\alpha \cdots  \beta_s \, i  \cdots j_f}\right)  &=  \sigma (1+2w)  \mathcal D_{\alpha i} W^{\alpha \cdots  \beta_s \, i \cdots j_f} 
                      +  (2w+ s) (\mathcal D_{\alpha i} \sigma)  W^{\alpha \cdots  \beta_s \, i \cdots j_f}  \nonumber \\                                                                                                                   
                                           &- 4(3 + s)  (\mathcal D_{\gamma i} \sigma)  W^{\gamma \cdots  \beta_s \, i \cdots j_f}   
				     +4(1+f )   (\mathcal D_{\alpha l } \sigma )   W^{\alpha \cdots  \beta_s \, l \cdots j_f}     \nonumber  \\
				     &+ 4   (\mathcal D_{\alpha l } \sigma)   W^{\alpha \cdots  \beta_s \, l \cdots j_f} ~.
\label{contraction1}
\end{align}
Therefore, if we require that the inhomogeneous parts of the  expression above cancel, necessarily the Weyl weight should be fixed in terms of the number of spin and isospin indices as
\begin{equation}
w = \tfrac{3}{2} s + 2(1-f) ~.
\label{wsf1}
\end{equation}
If so, the following constraint 
\begin{equation}
 \mathcal D_{\alpha i}  W^{(\alpha \cdots  \beta_s)\,( i  \cdots j_f) } = 0
\end{equation}
transforms homogeneously under Weyl transformations. In particular,  for the  spinor superfield $W^{\alpha k}$ ($s=1=f$) we can consider
 \begin{equation}
 \mathcal D_{k}  W^k = 0  \ ;\ \ \ \  w=\tfrac{3}{2} ~.
\label{sWconstraint1}
\end{equation}
Let us now consider the combination
$ (\tilde \gamma _{ab})^{\alpha}{}_{\beta_1}  \mathcal D_{\alpha( i} W^{\beta_1  \cdots  \beta_s}{}_ {k)}{} ^{j_2 \cdots j_f}$.
From (\ref{Wupper}) we get (not yet symmetrizing $_{(ik)}$)
\begin{align}
\varepsilon_{k j_1} (\tilde \gamma _{ab})^{\alpha}{}_{\beta_1} \delta \left( \mathcal D_{\alpha i}W^{\beta_1 \cdots  \beta_s \, j_1 \cdots j_f} \right)  &=
  \sigma (1+2w) (\tilde \gamma _{ab})^{\alpha}{}_{\beta_1}  \mathcal D_{\alpha i} W^{\beta_1  \cdots  \beta_s}{}_ {k}{} ^{j_2 \cdots j_f}\cr
                     &+  (2w+ s)(\tilde \gamma _{ab})^{\alpha}{}_{\beta_1} (\mathcal D_{\alpha i} \sigma)  W^{\beta_1  \cdots  \beta_s}{}_ {k}{} ^{j_2 \cdots j_f} \cr                                                                                                               
&-4  (\mathcal D_{\gamma i} \sigma)  \sum_{q=1}^{s}  (\tilde \gamma _{ab})^{\beta_q}{}_{\beta_1} W^{\beta_1 \cdots \beta_{q-1} \gamma \beta_{q+1} \cdots \beta_s }{}_{k}{}^{j_1 \cdots j_f}   \cr
&- 8 (\tilde \gamma _{ab})^{\alpha}{}_{\beta_1}  (\mathcal D_{\alpha}{}^{l} \sigma)  \sum_{q=1}^{f} \varepsilon_{k j_1}  \delta ^{j_q}_{(i} W^{\beta_1 \cdots \beta_s j_1\cdots  j_{q-1}} {}_{l)}{}^ {j_{q+1} \cdots j_f} ~. \cr
\label{constraintX}
\end{align}
We note that, in the first sum,  all the terms have the spin index of the covariant derivative and the superfield $W$ (the $\gamma$ index) contracted, except the term for $q=1$, which is identically zero.  This restrics us, for the transformation (\ref{constraintX}) to be homogeneous, to the case $s=1$.  The last sum (over the isospin indices), can be written as 
\begin{align}
\sum_{q=1}^{f} \varepsilon_{k j_1}  \delta ^{j_q}_{(i} W^{\beta_1 \cdots \beta_s j_1\cdots  j_{q-1}} {}_{l)}{}^ {j_{q+1} \cdots j_f}  &=
  \varepsilon_{k (i}  W^{\beta_1 \cdots \beta_s}{}_{l)}{}^{ j_2  \cdots j_f}  \nonumber \\ &+
   \sum_{q=2}^{f} \varepsilon_{k j_1}  \delta ^{j_q}_{(i} W^{\beta_1 \cdots \beta_s j_1\cdots  j_{q-1}} {}_{l)}{}^ {j_{q+1} \cdots j_f} ~,
\label{isosum}
\end{align}
and again, for the transformation to be homogeneous,  we need $f=1$ (otherwise there will be terms of the type $W_{\cdots i \cdots k \cdots}$ ). Then, taking the symmetric part in $_{(ij)}$ and combining (\ref{constraintX}) and (\ref{isosum}), we conclude that the only possibility for this kind of combination is necessarily the case $s=1=f$, that is
\begin{align}
\delta \left( \mathcal D_{\alpha(i} (\tilde \gamma _{ab})^{\alpha}{}_{\beta}  W^{\beta}{}_{j)} \right) &= \sigma (1+2w) \mathcal D_{\alpha(i} (\tilde \gamma _{ab})^{\alpha}{}_{\beta}  W^{\beta}{}_{j)} \nonumber \\
         &+  (2w+1)  (\mathcal D_{\alpha(i} \sigma )  (\tilde \gamma _{ab})^{\alpha}{}_{\beta}  W^{\beta}{}_{j)}  \nonumber \\
         &- 4 \mathcal D_{\alpha(j} (\tilde \gamma _{ab})^{\alpha}{}_{\beta}  W^{\beta}{}_{i)}   ~.
\label{}
\end{align}
Therefore, the only possible homogeneous  constraint is
\begin{equation}
\mathcal D_{(i} \tilde \gamma _{ab}  W_{j)} =0  ~,
\label{constraint2}
\end{equation}
with Weyl weight $w= \tfrac{3}{2}$. The abelian vector multiplet was precisely  described by this Weyl weight-$\tfrac{3}{2}$, spinor superfield $W^{\alpha i}$ subject to the constraints (\ref{sWconstraint1}) and (\ref{constraint2}). At the level of components, the first says that the vector multiplet auxiliary fields consist of an iso-triplet of scalars, while the second says that there is only one 2-form field strength (of the four possibles).
Together, these constraints can be written as
\begin{eqnarray}
\mathcal D_{\alpha i} W^{\beta}{}_{j} = \tfrac{1}{4} \, \delta_{\alpha}^{\beta}\, \mathcal D_{\gamma(i} W^{\gamma}{}_{j)}
  		-\tfrac{1}{16} \, \varepsilon_{ij}\, (\gamma^{ab})_{\alpha}{}^{\beta} \mathcal D_{k} \,\tilde \gamma_{ab}\, W^{k}  ~.
\end{eqnarray}
\section{Tensor multiplet}
Let $\tilde \Phi$ denote a six-dimensional real scalar superfield of Weyl-weight $w$, that is $\delta \tilde \Phi = 2 w \sigma \tilde \Phi$. Then, given such a transformation, its double spinorial derivative transforms into
\begin{align}
\delta (\mathcal D_{\alpha i}\mathcal D_{\beta j} \tilde \Phi) &= 
     2(1+w) \sigma \mathcal D_{\alpha i}\mathcal D_{\beta j} \tilde \Phi 
   +2w (\mathcal D_{\alpha i}\mathcal D_{\beta j} \sigma)  \tilde \Phi  \\
   &+2w (\mathcal D_{\alpha i}\sigma)  \mathcal D_{\beta j}  \tilde \Phi
   +4 (\mathcal D_{\beta i}\sigma) \mathcal D_{\alpha j}  \tilde \Phi
   -4 \,\varepsilon_{ij} (\mathcal D_{\alpha k}\sigma) \mathcal D_{\beta}{}^{k} \tilde \Phi \\
   &-2w(\mathcal D_{\beta j}\sigma) \mathcal D_{\alpha i}  \tilde \Phi
   -4 (\mathcal D_{\alpha j}\sigma) \mathcal D_{\beta i}  \tilde \Phi  ~.
\end{align}
Taking the symmetric part in isospin indices $_{(ij)}$ gives
\begin{align}
\delta (\mathcal D_{\alpha (i}\mathcal D_{\beta j)} \tilde \Phi)  =  2(1+w) \sigma \mathcal D_{\alpha (i}\mathcal D_{\beta j)} \tilde \Phi 
   &+2w (\mathcal D_{\alpha (i}\mathcal D_{\beta j)} \sigma) \tilde \Phi \nonumber\\
   &+4(w-2) (\mathcal D_{[\alpha (i}\sigma)  \mathcal D_{\beta] j)}  \tilde \Phi  ~.
\end{align}
In this last equation, the symmetric part in spin indices $_{(\alpha \beta)}$ is trivial. The appearance of the last term means that the anti-symmetric part can not be corrected to transform homogeneously unless $w=2$.  In this case, the second term can be cancelled by adding a connection term (\ref{sWC}). We define $\tilde \Phi = \Phi$ to be a real, weight-2, scalar supefield satisfying the invariant condition
\begin{eqnarray}
\mathcal D_{(i} \tilde \gamma_a \mathcal D_{j)} \Phi + 16i\, C_{\, ij}  \Phi =0~.
\label{Phicondition}
\end{eqnarray}
The same argument that lead to establishing the constraint in Eq. (\ref{sWconstraint1}) implies that if a superfield potential $V^{\alpha i}$ has weight $w=\tfrac{3}{2}$, the combination $\Phi:= \mathcal D_{\alpha i} V^{\alpha i}$ will be covariant under Weyl transformations and will have weight $w=2$. Then, it can be shown that, if $V^{\alpha i}$ satisfies only the constraint 
\begin{eqnarray}
\mathcal D_{(i} \tilde \gamma _{ab}  V_{j)} =0~,
\end{eqnarray}
then the associated scalar $\Phi$ satisfies the condition (\ref{Phicondition}). The scalar field $\Phi$ contains the \emph{anti}-self-dual field strength of a 2-form potential $H^{(-)}_{abc} \sim \mathcal D^{k}\tilde \gamma_{abc} \mathcal D_{k} \Phi |$. In terms of the potential superfield $V^{\alpha i}$, the potential  2-form is $B_{ab} \sim \mathcal D_{k} \tilde \gamma_{ab} V^{k}$. Thus, we find that the field $\Phi$ describes a real scalar, an anti-self-dual 3-form field strength, and their superpartners while the field $V^{\alpha i}$ describes the same multiplet in terms of a gauge 2-form potential.
\section{Other multiplets}
Let $\Phi_{\alpha(s) i(f)}:= \Phi_{\alpha_1 \cdots \alpha_s i_1 \cdots i_f}$ denote a superfield symmetric in $s$ spin and $f$ isospin indices. Let $w$ denote the Weyl-weight $\delta \Phi = 2w \sigma \Phi$. Then the combination $\mathcal D_{(\alpha( i} \Phi_{\beta(s)) i(f))}$, completely symmetrized on all indices, transforms homogeneously under Weyl transformations if and only if\footnote{We reach this result through a similar argument used to derive (\ref{wsf1}).}
\begin{eqnarray}
w= 2f - \tfrac{3}{2} s ~.
\end{eqnarray}
When this relation between Weyl-weight and spin and SU(2) indices is satisfied, the constraint 
\begin{eqnarray}
\mathcal D_{(\alpha( i} \Phi_{\beta(s)) i(f))}=0 ~,
\end{eqnarray}
can be imposed on the (matter) field $\Phi$. One can further confirm that this constraint is integrable in the sense that the anti-commutator 
$\{\mathcal D_{\alpha i}, \mathcal D_{\beta j} \} \Phi_{\gamma(s) k(f)}$ vanish identically when symmetrized on all spin and isospin indices.\footnote{Straightforward calculation shows that the isospin part will always be proportional to the SU(2) anti-symmetric tensor $\varepsilon_{ij}$, while the spin part will always reduce to having one $\gamma$-matrix. Therefore, after symmetrization, the anti-commutator in question vanish identically.} 
Examples include the hypermultiplet $q_i$ constrained by
\begin{eqnarray}
\label{hyper}
\mathcal D_{\alpha(i} \,q_{j)} =0  \ ;\ \ \ \ w=2~,
\end{eqnarray}
the isotriplet $L_{ij}$ subject to the constraint
\begin{eqnarray}
\mathcal D_{\alpha (i} L_{jk)} =0  \ ;\ \ \ \ w=4~,
\end{eqnarray}
the \textbf{5} of isospin $L_{ijkl}$ subject to the constraint
\begin{eqnarray}
\mathcal D_{\alpha (i} L_{jklm)} =0  \ ;\ \ \ \ w=8~,
\end{eqnarray} 
and the superfield $A_{\alpha i}$ subject to the constraint
\begin{eqnarray}
\mathcal D_{(\alpha (i} A_{\beta) j)} =0 \ ;\ \ \ \ w=\tfrac{1}{2}~.
\label{abelianFS}
\end{eqnarray}
Note that since $A_{\alpha i}$ has the same dimension as the covariant derivative $\mathcal D_{\alpha i}$, the replacement
$\mathcal D_{\alpha i} \rightarrow \nabla_{\alpha i} = \mathcal D_{\alpha i} + i A_{\alpha i}$ corresponds to minimal coupling to a super-1-form. It also follows that since $\{\mathcal D_{\alpha (i}, \mathcal D_{\beta j)} \} \, \Omega \equiv 0$ on a  scalar superfield $\Omega$, the constraint 
(\ref{abelianFS}) is invariant under the abelian gauge transformation $A_{\alpha i} \rightarrow A_{\alpha i} + \mathcal D_{\alpha i} \Omega$.

\section{Tensor calculus}
As anticipated at the beginning of this chapter, the constraints defining the scalar and tensor multiplets imply the on-shellness of these multiplets. In this section, we will compute for each of these matter representations the equations of motion arising from such constraints.
\paragraph {The scalar multiplet}  can be described by the iso-doublet, Weyl weight-2, scalar superfield $q^{\,i}$ subject to the constraint 
(\ref{hyper}) 
\begin{eqnarray}
\mathcal D_{\delta}{}^{(i}q^{\, j)}=0~.
\end{eqnarray}
A Weyl-type equation for $q^{\, i}$ arises by contracting the previous constraint with the operator 
$\varepsilon^{\alpha\beta\gamma\delta} \, \mathcal D_{\gamma i} D_{\delta j}$. Then, straightforward calculation leads to
\begin{eqnarray}
\label{Weylhyper}
\label{WeylScalar}
0= \mathcal D^{\alpha \beta} \mathcal D_{\beta i} q^{i} +2\, N^{\alpha\beta} \mathcal D_{\beta i} q^{i} 
 +48\, \mathcal C^{\alpha i} q_i - 10\, \mathcal N^{\alpha i} q_i ~.
\end{eqnarray}
Contracting again this equation with the spinorial covariant derivative, we obtain the scalar equation of motion
\begin{align}
\label{KGScalar}
0=\mathcal D^{a} \mathcal D_{a}q_i -\tfrac{17}{3}\, \mathcal D_{a} C^{a}{}_{ij} q^j &+2\, C^{a}{}_{ij} \mathcal D_{a} q^j
     + \tfrac{15}{2}\, C_{a jk} C^{a jk} q_i - 4N^{\alpha \beta}N_{\alpha \beta} q_i   \nonumber \\
     &-\tfrac{3i}{2}\, \mathcal D_{\alpha j} \mathcal C^{\alpha j} q_i -\tfrac{3i}{2}\, \mathcal C^{\alpha}{}_{i} \mathcal D_{\alpha j} q^j 
     + \tfrac{5i}{16} \, \mathcal D_{\alpha j} \mathcal N^{\alpha j} q_i ~,
\end{align}
where we have used the decompositions (\ref{DCdec}) and (\ref{DNdec}). Thess two equations of motion make manifest that the scalar multiplet is on-shell.

\paragraph{ The tensor muliplet} can be described by a real, Weyl weight-2, scalar superfield $\Phi$ subject to the constraint 
(\ref{Phicondition})
\begin{eqnarray}
\label{Phicondition2}
\mathcal D_{(i} \tilde \gamma_a \mathcal D_{j)}\Phi=-16i \,C_{a\, ij} \Phi ~.
\end{eqnarray}
Acting on this expression with the differential operator $(\mathcal D^j\tilde \gamma^a)^\alpha$ gives the spinor equation of motion
\begin{eqnarray}
\label{WeylTensor}
0= \mathcal D^{\alpha \beta} \mathcal D_{\beta}{}^{i} \Phi - C^{\alpha \beta ij} \mathcal D_{\beta j}\Phi 
    - 2\, N^{\alpha \beta} \mathcal D_{\beta}{}^{i} \Phi -12\, \mathcal C^{\alpha i} \Phi ~.
\end{eqnarray}
Contracting the equation above with a spinor covariant derivative results in a Klein-Gordon-type equation for the scalar $\Phi$\footnote{Alternatively, it can be shown that the same equation results from directly contracting the constraint (\ref{Phicondition2}) with the operator 
$\mathcal D^{(i} \tilde \gamma^{a} \mathcal D^{j)}$. }
\begin{equation}
\label{KGTensor}
0= \mathcal D^{a} \mathcal D_{a} \Phi + 8\, C_{a ij} C^{a ij} \Phi 
   + \tfrac{i}{6} \, N^{abc} \mathcal D^{k} \,\tilde \gamma_{abc} \mathcal D_{k} \Phi - \tfrac{3i}{2}\, \mathcal D_{\alpha i} \mathcal C_{\alpha i}
   -3i\, \mathcal C^{\alpha i} \mathcal D_{\alpha i} \Phi +\tfrac{5i}{2}\, \mathcal N^{\alpha i} \mathcal D_{\alpha i} \Phi .
\end{equation}
Again, we conclude that equations (\ref{WeylTensor}) and (\ref{KGTensor}) put the tensor multiplet automatically on-shell.

\chapter{Concluding remarks}
\label{ConcludingRemarks}
In this thesis, we have studied the geometry of six-dimensional, $\mathcal N=(1,0)$ superspace and its matter couplings.
In the first part of this work, we fixed the basic ground-work of our formulation by firstly setting the superspace structure group to $G=SO(5,1) \times SU(2)$. Then, after imposing the set of conventional torsion constraints (\ref{cc1}), (\ref{cc2}) and (\ref{cc2}), we systematically solved the supergravity Bianchi identities up to and including (mass) dimension-2. In doing so, we found that the full derivative algebra can  be expressed entirely in terms of a vector iso-triplet $C_{aij}$, and a 3-form  superfield $N_{abc}$. These superfields define the dimension-1 torsion, according to (\ref{dim1T}). We further elucidated that  consistency of the identities implies the constraints (\ref{constraint1})-(\ref{constraint3}) on these supergravity fields.  At dimension-$\frac{3}{2}$ we worked out the irreducible decomposition of torsion and isospin field strength. At the dimension-2 level, we computed the Riemann curvature tensor (\ref{RiemFinal}) and the field strength for R-symmetry group (\ref{FFinal}). 

Once we had in hand the complete supergeometry, we explored the invariance of the conventional torsion constraints under conformal transformations. In particular, we fixed the set of transformation rules that the superfields and covariant derivatives must satisfy in order to implement the conformal invariance. These transformations are given by (\ref{sWDspin}), (\ref{sWC}), (\ref{sWN}) and (\ref{sWDvector}). One of the important features of the transformation rules we found is that there are inhomogeneous pieces in the Weyl transformation of the superfields $C$ and $N$, which can be used to gauge away some of their components.

The second part of this thesis was dedicated to the study of the field content of the superspace supergravity presented in the first part, and its superconformal matter couplings. The Weyl multiplet was presented. As we have seen, its  gauge fields structure includes the frame field $e^{a}_{m}$, the gravitino $\psi^{\alpha i}_{m}$ and the (pure gauge) dilatation gauge field $B_{m}$. This set of gauge fields is encoded within the $\theta=0$ components of the superframe-field and superconnections. The matter field structure, on the other hand, is characterized by the set of fields $\{ N^{(-)}_{abc}, \chi^{\alpha i},F \}$, that is, an anti-self-dual tensor field, an auxiliary spinor and a real auxiliary scalar, respectively. These fields arise from the $\theta=0$ components of the three-form superfield $N_{abc}$ and its spinorial covariant derivative(s), as indicated in (\ref{theta0type1}).

Next, we investigated the possible matter fields allowed by conformal invariance. We started  by addressing the question of what are the most general conformally invariant constraints on a certain matter superfields. We then used those constraints to further study the (abelian) vector and tensor multiplets. The former turns out to be described by a Weyl-weight-$\frac{3}{2}$ spinor superfield $W^{\alpha i}$ subject to the constraints (\ref{sWconstraint1}) and (\ref{constraint2}), while the latter is characterized by a real, weight-2 scalar superfield $\Phi$ satisfying the condition (\ref{Phicondition}). This scalar field
admits a weight-$\frac{3}{2}$ potential $V$, defined through $\Phi= \mathcal D_{\alpha i} V^{\alpha i}$, which allows an alternative description of the same tensor multiplet.

We concluded this thesis with the study of the component field equations of motion for the scalar and tensor multiplet.
It was shown that, starting with the constraints defining a matter representation, one may further derive Weyl-type and Klein-Gordon-type equations of motion for the component fields defining each multiplet. These equations are given by
(\ref{WeylScalar}), (\ref{KGScalar}), (\ref{WeylTensor}) and (\ref{KGTensor}), and imply that both multiplets are realized on-shell.

Summarizing, this thesis may be considered as a companion to reference \cite{Linch:2012zh}, developing the very first steps and basic results in order to carry out further and deeper explorations of simple six-dimensional superspaces and their applications. There remains, therefore, much work to be done. A natural direction for future research is the dimensional reduction to five dimensions, with the hope of recovering the five-dimensional superfield supergravity presented in \cite{Kuzenko:2007hu}. It would also be desirable, since simple six-dimensional supergravity enjoys the same fermionic structure that of four-dimensional, $\mathcal N=2$ supergravity, to address the issue of how the latter is embedded in six-dimensional superspace.

The study of supersymmetric backgrounds in superspace is also a open problem. Along this line, one might attempt to extend early classifications of the geometries admissible for a six-dimensional supergravity description \cite{Gutowski:2003rg} to superspace. More ambitiously, the extension of lower-dimensional rigid supersymmetric backgrounds \cite{Festuccia:2011ws, Dumitrescu:2012ha, Cassani:2012ri} to six-dimensional curved superspace may be investigated. 

\appendix

\chapter{6D notation and conventions}
\label{appendixA}
We adopt the 6D superspace conventions established in \cite{Linch:2012zh}. The procedure is to first define $\gamma_{m}:= - \Gamma_{m} C^{-1}$ and $\tilde \gamma_{m} = - C\Gamma_{m}$ for $m=0,\dots, 3;5$. Then we take $\gamma_6 = C^{-1}$ and $\tilde \gamma_6 = -C$. The relative sign has been chosen so that the six $8\times 8$ Dirac matrices satisfy the Clifford algebra
\begin{align}
\left\{ \Gamma_{m}, \Gamma_{n} \right\}  = -2 \eta_{mn} \mathbf 1  ~,
\end{align}
with $m, n =0 , \dots, 5$
and
\begin{equation}
\eta_{mn}=\text{diag}(-1,1,1,1,1,1) ~.
\end{equation}
The overall sign is chosen so that, in terms of explicit indices, the formul\ae{}   are 
\begin{align}
(\gamma^a)_{\alpha \beta} = (\Gamma^a)_{\alpha \beta}
&,~~
(\tilde \gamma^a)^{\alpha \beta} = - (\Gamma^a)^{\alpha \beta}
~~~\mathrm{for}~ a=0,1,2,3;5\cr
(\gamma_6)_{\alpha \beta} = \varepsilon_{\alpha \beta} &,~~(\tilde \gamma_6)^{\alpha \beta}=-\varepsilon^{\alpha \beta}~.
\end{align}
In terms of Pauli-type matrices, Dirac matrices take the form
\begin{align}
\Gamma_{m}  = \left(
\begin{array}{cc}
0& (\gamma_{m})_{\alpha \beta}\\ 
(\tilde \gamma_{m})^{\beta \alpha}&0
\end{array}
\right) ~,
\end{align}
with $\alpha = 1,\dots,4$. 

It is possible to represent these 6D Pauli-type matrices $\gamma_m$ and $\tilde\gamma_m$,
in terms of the 4D Pauli matrices, $\sigma_m$ and $\tilde\sigma_m$. Denoting the 4D, SL$(2,\mathbb{C})$ spinor indices by 
$\alpha=1,2$ and $\dot\alpha=1,2$, such representation is given by
\begin{align}
\label{GammaMap1}
 \gamma_m = \left(
\begin{array}{cc}
0&- (\sigma_m)_\alpha{}^{\dot \beta}\\ 
(\tilde \sigma_m)^{\dot \alpha}{}_{ \beta}&0
\end{array}
\right)
\end{align}
for $m=0,\dots,3$ and 
\begin{align}
\label{GammaMap2}
\left( \gamma_5\right)_{\alpha \beta} = \left(
\begin{array}{cc}
i \varepsilon_{\alpha \beta}&0\\ 
0& i \varepsilon^{\dot \alpha \dot \beta}
\end{array}
\right)
,~~~
\left( \gamma_6\right)_{\alpha \beta} = \left(
\begin{array}{cc}
\varepsilon_{\alpha \beta}&0\\ 
0& - \varepsilon^{\dot \alpha \dot \beta}
\end{array}
\right).
\end{align}
Defining now
\begin{align}
\label{GammaMap2}
\left(\tilde \gamma_5\right)^{\alpha \beta} = \left(
\begin{array}{cc}
i \varepsilon^{\alpha \beta}&0\\ 
0& i \varepsilon_{\dot \alpha \dot \beta}
\end{array}
\right)
,~~~
\left(\tilde \gamma_6\right)^{\alpha \beta} = \left(
\begin{array}{cc}
-\varepsilon^{\alpha \beta}&0\\ 
0&  \varepsilon_{\dot \alpha \dot \beta}
\end{array}
\right) ~,
\end{align}
six-dimensional Pauli-type matrices obey the algebra
\begin{align}
(\gamma^{m})_{\alpha \beta}( \tilde \gamma^{n})^{\beta \gamma}
	+(\gamma^{n})_{\alpha \beta}(\tilde \gamma^{m})^{\beta \gamma} 
	= -2 \eta^{mn} \delta_{\alpha}^{\gamma}  ~,
\cr
( \tilde \gamma^{m})^{\alpha \beta}(\gamma^{n})_{\beta \gamma}
	+( \tilde \gamma^{n})^{\alpha \beta}(\gamma^{m})_{\beta \gamma} 
	= -2 \eta^{mn} \delta^{\alpha}_{\gamma}  ~.
\end{align}
Note that the 6-dimensional Pauli-type matrices are antisymmetric
\begin{align}
(\gamma_{m})_{\alpha \beta}  = -(\gamma_{m})_{\beta \alpha} ,
\end{align}
implying an isomorphism between the space of 6-dimensional vectors and antisymmetric $4\times 4$ matrices
\begin{align}
V_{\alpha \beta}:= (\gamma^{m})_{{}\alpha {}\beta}V_{m}=-V_{\beta \alpha}
~~~\Leftrightarrow~~~ V_{m} =  \tfrac14 (\tilde \gamma_{m})^{\alpha \beta} V_{\alpha \beta}  ~.
\end{align}
The second relation is a consequence of the analysis below and equation (\ref{6Fierz2}) in particular.
Similarly, six-dimensional 2-forms are in one-to-one correspondence with traceless $4\times 4$ matrices and (anti-)self-dual 3-forms are in correspondence with symmetric rank-2 spin matrices with their indices (up) down as we now work out in detail. 
To begin, it is useful to define the normalized anti-symmetrized products of Pauli-type matrices
\begin{align}
{\gamma}_{m_1\dots m_p} &:= \gamma_{[m_1} \tilde \gamma_{m_2}\dots 
	\stackrel{\mbox{\tiny{$(\sim)$}}}{\gamma}\!\!\!\!{}_{m_p]} 
=  \tfrac1{p!} \gamma_{m_1} \tilde \gamma_{m_2}\dots \stackrel{\mbox{\tiny{$(\sim)$}}}{\gamma}\!\!\!\!{}_{m_p} + \mathrm{perm.}\cr
\tilde {\gamma}_{m_1\dots m_p} &:= \tilde \gamma_{[m_1} \gamma_{m_2}\dots \stackrel{\mbox{\tiny{$(\sim)$}}}{\gamma}\!\!\!\!{}_{m_p]} =  \tfrac1{p!} \tilde \gamma_{m_1} \gamma_{m_2}\dots \stackrel{\mbox{\tiny{$(\sim)$}}}{\gamma}\!\!\!\!{}_{m_p} + \mathrm{perm.}
\end{align}
With these normalizations products reduce without factors. For example 
\begin{align}
{\gamma}^{ab} \gamma^c = {\gamma}^{abc} +2\eta^{c[a}\gamma^{b]} 
~~~,~~~
\tilde \gamma^c {\gamma}^{ab}  = \tilde {\gamma}^{abc} -2\eta^{c[a}\tilde \gamma^{b]}  ~.
\end{align}
Other useful identities are 
\begin{align}
\label{g3g1}
\gamma_{abd}\tilde \gamma_c &= - \tfrac12 \epsilon_{abdcef}\gamma^{ef} -3 \eta_{c[a}\gamma_{bd]}~,  \\
\tilde \gamma_{abd} \gamma_c &=
	+ \tfrac12 \epsilon_{abdcef}\tilde\gamma^{ef} 
	-3 \eta_{c[a}\tilde\gamma_{bd]}~,	 \\
\tilde \gamma^c \gamma_{gh} \gamma_f &=
	 \tfrac 12 \epsilon^c{}_{ghfrs}\tilde \gamma^{rs} 
	-3 \delta^c_{[g}\tilde \gamma_{hf]}
	-2\eta_{f[g}\tilde \gamma_{h]}{}^c
	+ 2\delta^c_{[g}\eta_{h]f}~,  \\
\tilde \gamma^d\gamma_{abc}\tilde \gamma_d&=0~, \\
\gamma^d\tilde \gamma_{abc}\gamma_d&=0~, \\
{\gamma}^{ab} \gamma_{cd} &= 
	- \tfrac12 \epsilon^{ab}{}_{cdef}\gamma^{ef} 
	+4 \delta^{[a}_{[c} \gamma^{b]}{}_{d]}
	-2 \delta^{[a}_{[c}\delta^{b]}_{d]} ~,\\
\label{32}
{\tilde \gamma}_{abc} \gamma_{de} &= 
	 \tfrac12 \epsilon_{abcd}{}^{fg}\tilde \gamma_{efg}
	+\epsilon_{abcdef}\tilde \gamma^f
	+\eta_{de}\tilde \gamma_{abc}
	-3\eta_{d[a}\tilde \gamma_{bc]e}
	+6\eta_{d[a}\tilde \gamma_{b}\eta_{c]e}~.~~~~~~
\end{align}	
A more commonly used convention regarding the 2-form matrix is as the spinor representation (\ref{LRep}) of the Lorentz generator $M_{ab}$ which is related by
\begin{align}
\label{SpinNorm}
(\Sigma^{ab})_\alpha{}^\beta = -  \tfrac12 (\gamma^{ab})_\alpha{}^\beta ~.
\end{align}
In terms of these matrices, we define
\begin{align}
F_{\alpha}{}^{\beta}:=  \tfrac12 ({\Sigma}^{m n})_{{}\alpha}{}^{{}\beta}F_{mn}
~~\Rightarrow~~  F_{mn}= - ({\Sigma}_{m n})_{{}\beta}{}^{{}\alpha} F_{\alpha}{}^{\beta}~.
\end{align}
The second relation is a consequence of the analysis below and equation (\ref{Fierz2form}) in particular. Both equations again agree with the five-dimensional conventions. Using the second type of matrix, we can construct $\tilde{F}^{\alpha}{}_{\beta}:= \tfrac14 (\tilde{{\gamma}}^{m n})^{{}\alpha}{}_{{}\beta}F_{mn}$, however 
\begin{align}
(\tilde {\gamma}^{mn})^{\alpha}{}_{\beta} = - ({\gamma}^{mn})_{\beta}{}^{\alpha} ~,
\end{align}
so that this second matrix is not essentially new.
Finally, the third-rank antisymmetric tensors can be separated into (anti-)self-dual parts which are then in one-to-one correspondence with symmetric $4\times4$ matrices. To see how this works in detail, we must first establish some Fierz identities. There is a completeness relation
\begin{align}
\label{6Fierz1}
 \tfrac12 (\gamma^{m})_{\alpha \beta} (\gamma_{m})_{\gamma \delta} = \varepsilon_{\alpha \beta \gamma \delta} ~.
\end{align}
Contraction with $\varepsilon^{\gamma^\prime \delta^\prime \gamma \delta}$ implies the completeness relation
\begin{align}
\label{6Fierz2}
 \tfrac12 (\gamma^{m})_{\alpha \beta} (\tilde \gamma_{m})^{\gamma \delta} = \delta_{\alpha}^{ \gamma} \delta_{ \beta}^{ \delta} -\delta_{\beta}^{\gamma} \delta_{ \alpha}^{ \delta}~,
\end{align}
and\footnote{This relation follows, up to normalization, from the equal dimensions of the spaces of 6-vectors and antisymmetric $4\times4$ matrices.} 
\begin{align}
 \tfrac12 \varepsilon^{\alpha \beta \gamma \delta} (\gamma_{m})_{\gamma \delta} =  (\tilde \gamma_{m})^{\alpha \beta}
~~\Rightarrow~~
(\gamma_{m})_{\alpha \beta} = \tfrac12 \varepsilon_{\alpha \beta \gamma \delta}   (\tilde \gamma_{m})^{\gamma \delta}~.
\end{align}
Contraction of (\ref{6Fierz2}) with itself gives
\begin{align}
\label{Fierz2form}
 \tfrac 14 (\tilde {\gamma}^{mn})^{\alpha}{}_{\beta}({\gamma}_{mn})_{\gamma}{}^{\delta} = - \tfrac12 \delta_{\beta}^{\alpha}\delta_{\gamma}^{\delta} +2 \delta_{\beta}^{\delta}\delta_{\gamma}^{\alpha}~.
\end{align}
Another contraction with (\ref{6Fierz2}) gives
\begin{align}
\label{Fierz3form}
(\tilde {\gamma}^{abc})^{\alpha \beta} ({\gamma}_{abc})_{\gamma \delta} =24(\delta^{\alpha}_{\gamma}\delta^{\beta}_{\delta} + \delta^{\alpha}_{\delta}\delta^{\beta}_{\gamma})~,
\end{align}
while contraction with (\ref{6Fierz1}) shows that 
\begin{align}
({\gamma}^{abc})_{\alpha \beta} ({\gamma}_{abc})_{\gamma \delta} = 0 
	&~~\mathrm{and}~~ 
		(\tilde {\gamma}^{abc})^{\alpha \beta}( \tilde {\gamma}_{abc})^{\gamma \delta} = 0~.
\end{align}
Thus we see that $\tilde {\gamma}^{mnp}$ and ${\gamma}^{mnp}$ correspond to (anti-)self-dual 3-forms. To show that ($\tilde {\gamma}^{mnp}$) ${\gamma}^{mnp}$ is (A)SD, one uses the identities 
\begin{align}
\gamma_0\tilde \gamma_1\gamma_2\tilde \gamma_3\gamma_5\tilde \gamma_6 = +\mathbf 1
	&~~\mathrm{and}~~ 
\tilde \gamma_0\gamma_1\tilde \gamma_2\gamma_3\tilde \gamma_5 \gamma_6 = -\mathbf 1	
\end{align}
to conclude that, for example, $\gamma_{012} = \epsilon_{012}{}^{345} \gamma_{345}$ whereas $\tilde \gamma_{012} = - \epsilon_{012}{}^{345} \tilde \gamma_{345}$.
The relation (\ref{32}) immediately implies the trace relation on the 3-forms
\begin{align}
\mathrm {tr} (\tilde \gamma_{abc}\gamma^{def}) = 
	4! \left( \delta_{[a}^d\delta_b^e\delta_{c]}^f - \tfrac1{3!} \epsilon_{abc}{}^{def} \right) ~,
\end{align}
from which it follows that the (anti-)self-dual parts of a 3-form $N$ satisfy
\begin{align}
\slash\!\!\!\! N^{(+)\alpha \beta}&:=  \tfrac1{3!} N_{abc} (\tilde \gamma^{abc})^{\alpha \beta} 
	~~\Rightarrow~~N^{(+)}_{abc}= \tfrac18 \mathrm {tr}(\,\slash\!\!\!\! N^{(+)} \gamma_{abc})~, \cr
\slash\!\!\!\! N^{(-)}_{\alpha \beta}&:=  \tfrac1{3!} N_{abc} (\gamma^{abc})_{\alpha \beta} 
	~~\Rightarrow~~  N^{(-)}_{abc}= \tfrac18 \mathrm {tr}(\,\slash\!\!\!\! N^{(-)} \tilde\gamma_{abc}) ~.
\end{align}
Recall that (six-dimensional) Hodge duality on 3-forms is an involution of order 2:
\begin{align}
 \tfrac 1{3!} \epsilon_{abcrst}\epsilon^{defrst} = - 3! \delta_{[a}^d\delta_b^e \delta_{c]}^f ~.
\end{align}
Other useful relations resulting from (\ref{6Fierz1}) are
\begin{align}
\label{6FierzHigh1}
(\gamma_a)_{\alpha \beta} ({\gamma}^{ab})_\gamma{}^\delta 
	&=  2 \varepsilon_{\alpha \beta \gamma \epsilon} (\tilde \gamma^b)^{\epsilon \delta} + (\gamma^b)_{\alpha \beta} \delta_\gamma^\delta~,\\
\label{6FierzHigh2}
(\gamma_a)_{\alpha \beta} ({\gamma}^{abc})_{\gamma \delta}
	&=  -2 \varepsilon_{\alpha \beta \gamma \epsilon} ({\gamma}^{bc})_\delta{}^\epsilon + 2(\gamma^{[b})_{\alpha \beta} (\gamma^{c]})_{\gamma \delta	}~,\\
\label{6FierzHigh3}
({\gamma}_{abc})_{\alpha \beta} ({\gamma}^{bc})_\gamma{}^\delta &= - 8 (\gamma_a)_{\gamma(\alpha}\delta_{\beta)}^\delta~.
\end{align}
Further contractions of these equations give a long list of useful identities, namely
\begin{align}
\label{3c1}
(\gamma_{abc})_{\alpha \beta} (\tilde \gamma^c)^{\gamma \delta} &= -4 \delta_{(\alpha}^{[\gamma} (\gamma_{ab})_{\beta)}{}^{\delta]}~,\\ 
\label{3c3}
(\gamma_{acd})_{\alpha \beta} (\tilde \gamma_b{}^{cd})^{\gamma \delta} &= 8\eta_{ab}\delta_{(\alpha}^{\gamma} \delta_{\beta)}^\delta
-8 \delta_{(\alpha}^{(\gamma} (\gamma_{ab})_{\beta)}{}^{\delta)}~,\\ 
(\gamma_{abc})_{\alpha \beta} (\gamma^c)_{\gamma \delta} &= -4 (\gamma_{[a})_{\gamma(\alpha}(\gamma_{b]})_{\beta)\delta}~,\\ 
\label{3c3SD}
(\gamma_{acd})_{\alpha \beta} (\gamma_b{}^{cd})_{\gamma \delta} &= -8 (\gamma_{(a})_{\gamma(\alpha}(\gamma_{b)})_{\beta)\delta}~,\\ 
(\tilde \gamma_{abc})^{\alpha \beta} (\gamma^c)_{\gamma \delta} &= 4 
\delta_{[\gamma}^{(\alpha} (\gamma_{ab})_{\delta]}{}^{\beta)}~,\\
\label{3c2}
(\tilde \gamma_{abc})^{\alpha \beta} (\gamma^{bc})_\gamma{}^\delta &= 8(\tilde \gamma_a)^{\delta(\alpha}\delta_\gamma^{\beta)}~.
\end{align}
Let us conclude deriving some other useful relations. Starting from the second relation above and contracting with $\varepsilon^{\mu \nu \gamma \delta}$ gives
\begin{align}
0=\delta_{[\alpha}^\mu ({\gamma}^{bc})_{\beta]}{}^\nu -\delta_{[\alpha}^\nu ({\gamma}^{bc})_{\beta]}{}^\mu+(\gamma^{[b})_{\alpha \beta}(\tilde \gamma^{c]})^{\mu \nu}
\end{align}
Next, contraction with $(\gamma^a)_{\gamma \mu}$ gives
\begin{align}
({\gamma}^{bc} \gamma^a)_{[\alpha|\gamma|}\delta_{\beta]}^\nu&=(\gamma^{[b})_{\alpha \beta}(\gamma^{|a|}\tilde \gamma^{c]})_\gamma{}^\nu+(\gamma^a)_{\gamma[\alpha} ({\gamma}^{bc})_{\beta]}{}^\nu~.
\end{align}
Taking the completely antisymmetric part $[abc]$ gives the identity
\begin{align}
\label{3gamma}
({\gamma}^{abc})_{\gamma [ \alpha} \delta_{\beta]}^\delta = 
	- (\gamma^{[a})_{\alpha \beta} ({\gamma}^{bc]})_\gamma{}^\delta 
	+(\gamma^{[a})_{\gamma[\alpha} ({\gamma}^{bc]})_{ \beta]}{}^\delta ~.
\end{align}
Finally, we can use the fact that
\begin{equation}
\varepsilon_{\alpha \beta \gamma \delta} \varepsilon^{\mu \nu \lambda \delta} = 
	3! \, \delta_{[\alpha}^\mu \delta_{\beta}^\nu \delta_{\gamma]}^\lambda ~,
\end{equation}
to show that 
\begin{align}
(\gamma^a)_{\gamma[\alpha } \psi_{\beta]} = 
	 \tfrac 14 \varepsilon_{\alpha \beta \gamma \delta} \varepsilon^{\mu \nu \lambda \delta} (\gamma^a)_{\mu \nu} \psi_{\lambda}- \tfrac12 (\gamma^a)_{\alpha \beta} \psi_{\gamma} ~,
\end{align}
and therefore
\begin{align}
\label{BigFierz}
({\gamma}^{abc})_{\gamma [ \alpha} \delta_{\beta]}^\delta = 
	- \tfrac32 (\gamma^{[a})_{\alpha \beta} ({\gamma}^{bc]})_\gamma{}^\delta 
	+ \tfrac14 \varepsilon_{\alpha \beta \gamma \rho} \varepsilon^{\mu \nu \lambda \rho}
		(\gamma^{[a})_{\mu \nu} ({\gamma}^{bc]})_{ \lambda}{}^\delta ~.
\end{align}


\chapter{Supergeometry summary}
\label{appendixB}

\paragraph{Superalgebra}  The dimension-1 and -$\tfrac{3}{2}$ commutators are given by
\begin{align} 
\{ \mathcal D_{\alpha i}, \mathcal D_{\beta j} \} &= 2i\, \varepsilon_{ij} (\gamma^a)_{\alpha \beta} \mathcal D_a
+2i\, (\gamma^{abc})_{\alpha \beta} C_{a ij} M_{bc} +4i\, \varepsilon_{ij} (\gamma_{a})_{\alpha \beta} N^{abc}M_{bc} \nonumber \\
&-6i\, \varepsilon_{ij} (\gamma^{a})_{\alpha \beta} C_{a}{}^{kl} J_{kl}-\tfrac{8i}{3} \, (\gamma^{abc})_{\alpha \beta} N_{abc} J_{ij} \\
\left[ \mathcal D_{\gamma k} , \mathcal D_{a}\right] &= -C^{b}{}_{kl} (\gamma_{ab})_{\gamma}{}^{\delta} \mathcal D_{\delta}{}^{l} 
   									+ N_{abc} (\gamma^{bc})_{\gamma}{}^{\delta} \mathcal D_{\delta k} + 
i \left[ \tfrac{1}{2} (\gamma_a)_{\gamma \delta}T_{bc}{}^{\delta}{}_k
       -(\gamma_{[b})_{\gamma \delta} T_{c]a}{}^{\delta}{}_k\right]M^{bc}   \nonumber\\
       & - \left[ (\gamma_a)_{\gamma \delta} \mathcal C^{\delta\,ij}{}_{k}  - 6\,\delta^{(i}_k \mathcal C_{a\,\gamma}{}^{j)}
    +5\,\delta^{(i}_k (\gamma_a)_{\gamma \delta}\left( \mathcal C^{\delta j)} -\tfrac13 \mathcal N^{\delta j)}\right)  \right] J_{ij}     
\end{align}

\paragraph{Irreducibles} Spinorial derivatives of the supergravity fields decompose as 
\begin{align}
\mathcal D_{\gamma k} C_{a\, ij} &= \mathcal C_{a \, \gamma k \, ij} + (\gamma_{a})_{\gamma \delta} \, \mathcal C^{\delta}{}_{ijk} 
            + \varepsilon_{k(i} \,\mathcal C_{a \gamma j)}+ \varepsilon_{k(i}(\gamma_{a})_{\gamma \delta} \, \mathcal C^{\delta}{}_{j)} \\
\mathcal D_{\gamma k} N_{\alpha \beta} &=
  \mathcal N_{\gamma k \, \alpha \beta} + \check{\mathcal N}_{\gamma k \, \alpha \beta}  \\ 
\mathcal D_{\gamma k} N^{\alpha \beta} &= \mathcal N_{\gamma k }{}^{ \alpha \beta} 
  						+ \delta_{\gamma}^{(\alpha} \mathcal N^{\beta)}{}_{k}
 \end{align}
 Under this decomposition, dimension-1 torsion constraints are equivalent to
 \begin{eqnarray}
\begin{array}{lcl}
\mathcal C_{a\, \gamma k\, ij}  = 0 &~~~& \mathcal N_{\gamma k\, \alpha \beta }=0\\ 
\mathcal C^{\delta}{}_{ijk}=- \tfrac{1}{6} (\tilde \gamma^b)^{\delta \beta} \mathcal D_{\beta(k} C_{b\, ij)}
	&&\check{\mathcal N}_{\gamma k \, \alpha \beta}=-  \tfrac{3}{4} (\gamma^a)_{\gamma(\alpha} \mathcal C_{a\,\beta)k}\\ 
\mathcal C_{a\, \beta j}=  \tfrac{1}{9}\tau_{a\, \beta}^{c\, \gamma} \mbox{\footnotesize{$(5,1)$}} \mathcal D_{\gamma}{}^{i} C_{c\, ij}
	&&\mathcal N_{\gamma k}{}^{\alpha \beta} = \mathcal D_{\gamma k} N^{\alpha \beta} 
	- \tfrac{2}{5} \delta_\gamma^{(\alpha} \mathcal D_{\delta k}  N^{\beta)\delta}\\ 
\mathcal C^{\gamma k}= - \tfrac{1}{9} \mathcal D_{\delta l} C^{\delta\gamma\, lk} 
	&&\mathcal N^{\alpha i} =  \tfrac{2}{5} \mathcal D_{\beta}{}^{i}  N^{\beta\alpha}
\end{array}	
\end{eqnarray}
The irreducible parts of the dimension-$\tfrac{3}{2}$ torsion and isospin field strength are
\begin{align}
T_{ab}{}^{\gamma k}&=\mathfrak T_{ab}{}^{\gamma k} + (\tilde \gamma_{[a})^{\gamma \delta}\, \mathfrak T_{b] \delta}{}^{k}
                                       + (\gamma_{ab})_{\delta}{}^{\gamma} \, \mathfrak T^{\delta k}   \\
F_{a\, \gamma k}{}^{ij}&= \mathfrak{F}_{a\, \gamma k}{}^{ij} + (\gamma_a)_{\gamma \delta} \, \mathfrak{F}^{\delta}{}_{k}{}^{ij}
             			      + \delta_{k}^{(i} \,\mathfrak{F}_{a\, \gamma}{}^{j)}  
			               + \delta_{k}^{(i} \, (\gamma_a)_{\gamma \delta} \, \mathfrak{F}^{\delta i)}
\end{align}
where
 \begin{eqnarray}
\begin{array}{lcl}
\mathfrak T_{ab}{}^{\gamma k} =
   -\tfrac{i}{2}\, (\gamma_{ab})_{\beta}{}^{\delta} \mathcal N_{\delta}{}^{k\, \beta \gamma}
   &~~~~&  \mathfrak{F}_{a\, \gamma k \, ij} =0 \\ 
\mathfrak T_{a\, \beta j}= -\tfrac{7i}{4} \, \mathcal C_{a\, \beta j}
	&& \mathfrak{F}^{\delta}{}_{k\, ij} =- \mathcal C^{\delta}{}_{ijk}\\ 
\mathfrak T^{\delta k} =   -i\, \mathcal C^{\delta k} + \tfrac{i}{6}\, \mathcal N^{\delta k}
	&& \mathfrak{F}_{a\, \gamma k} =  6\, \mathcal C_{a\, \gamma k} \\ 
	&& \mathfrak{F}^{\alpha i} = -5\, \mathcal C^{\alpha i} + \tfrac{5}{3} \, \mathcal N^{\alpha i} 
\end{array}	
\end{eqnarray}

\paragraph{Riemann and Ricci tensors, Curvature Scalar, SU(2) Field Strength}
At dimension-2 level, Bianchi identities encode the Riemann tensor
\begin{align}
R_{ab}{}^{cd} &=
\tfrac{i}{8} (\gamma^{cd})_{\beta}{}^{\alpha} (\gamma_{ab})_{\gamma}{}^{\delta} \mathbf{N}_{\alpha \delta}{}^{\beta\gamma}
+2\, \varepsilon_{ab}{}^{cdmn} \mathcal D^p \left[ N^{(+)}_{mnp} - \tfrac{4}{3}\, N^{(-)}_{mnp}\right]  \nonumber \\
&+ 4\, \mathcal D_{[a} N_{b]}{}^{cd} + 4\, \mathcal D^{[c} N^{d]}{}_{ab}  
-32\, N_{e[a}{}^{[c} N^{d]}{}_{b]}{}^{e} + 8\, \delta_{[a}^{[c} C_{b] ij}C^{d] ij} \nonumber  \\
&+ \tfrac{i}{2} \, \delta_{[a}^{[c} \delta_{b]}^{d]} \left[ \mathcal D_{\alpha i} \mathcal C^{\alpha i} + 8i\, C_{n ij} C^{n ij}
 -\tfrac{1}{6}\, \mathcal D_{\alpha i} \mathcal N^{\alpha i}   \right]  
\end{align}
It also follows that
\begin{align}
R_{ab} &=\tfrac{i}{8}  \eta_{ab} \left[ 10\: \mathcal D_{\alpha i} \mathcal C^{\alpha i} - \tfrac{5}{3}\: \mathcal D_{\alpha i} \mathcal N^{\alpha i} +64i \: C^{dij} C_{dij}  \right]
                                 + 8 \: C_{a}{}^{ij}  C_{bij}   
             +  16\, N^{cd}{}_{a} N_{bcd~~}     \\
R&= \tfrac{15i}{2}\: \mathcal D_{\alpha i} \mathcal C^{\alpha i}  - 40 \: C_{aij}C^{aij}
           -\tfrac{5i}{4}\: \mathcal D_{\alpha i} \mathcal N^{\alpha i} + 16\: N_{abc} N^{abc}  
\end{align}
Finally, the dimension-2 SU(2) field strength is given by
\begin{align}
F_{ab}{}^{ij} &= \tfrac{5i}{12}\, \mathbf{N}_{ab}{}^{ij} - \tfrac{11i}{288} \, \mathbf C_{ab}{}^{ij} +\tfrac{5i}{18} \, \mathbf{\tilde C}_{ab}{}^{ij}  
   + \tfrac{10}{9} \, \mathcal D_{[a} C_{b]}{}^{ij} \nonumber \\
   &+ \tfrac{86}{9}\, C_{[a}{}^{k(i} C_{b]}{}^{j)}{}_{k} 
   + \tfrac{4}{9}\,N^{(+)}_{abd} C^{dij} + \tfrac{200}{9}\,N^{(-)}_{abd} C^{dij}
\end{align}

\paragraph{Super-Weyl transformations} Covariant derivatives and superfields transform as
\begin{align}
\delta \mathcal D_{\alpha i} &= \sigma \mathcal D_{\alpha i} -4 \, (\mathcal D_{\beta j} \sigma) M_{\alpha}{}^{\beta}   
 										           +8 \, (\mathcal D_{\alpha}{}^{j} \sigma) J_{ij} \\
\delta \mathcal D_a &= 2\, \sigma \, \mathcal D_a -i\, (\mathcal D^k \sigma)\, \tilde \gamma_a \, \mathcal D_k
    - 2\, (\mathcal D^b \sigma)\, M_{ab} - \tfrac{i}{4} \, (\mathcal D^{i}\tilde \gamma_{a} \mathcal D{j}  \sigma)  J_{ij}   	\\	
\delta C_{aij} &= 2\sigma C_{aij}  + \tfrac{i}{4} \, \mathcal D_{(i}\tilde \gamma_{a} \mathcal D_{j)} \sigma  \\
\delta N_{abc} &= 2\sigma N_{abc} - \tfrac{i}{16}\, \mathcal D^{k} \tilde \gamma_{abc} \mathcal D_{k} \, \sigma   
\end{align}


\end{document}